\documentclass[prl,showkeys,showpacs,footnoteinbib]{revtex4-2}
\usepackage[utf8]{inputenc}
\usepackage{amsmath} 
\usepackage{amsfonts}
\usepackage{amssymb}
\usepackage{epsfig}
\usepackage{siunitx}
\usepackage{todonotes}

\begin{document}

\title{Fluctuation dissipation relations for strongly correlated out-of-equilibrium circuits}

\author{In\`es Safi}

\address{Laboratoire de Physique des Solides-CNRS-UMR $5802$. University Paris-Saclay, B\^at. $510$, $91405$ Orsay, France}
\begin{abstract}
We consider strongly correlated quantum circuits where a dc drive is added on top of an initial
out-of-equilibrium (OE) stationary state. Within a perturbative approach, we
derive unifying OE fluctuation relations
for high frequency current noise, shown to be completely determined by
zero-frequency noise and dc current. We apply them to the fractional
quantum Hall effect at arbitrary incompressible filling factors, driven by OE sources,
without knowledge of the underlying model. We show that such OE relations provide robust methods for
an unambiguous determination of the fractional charge 
or of key interaction parameters entering in the
exploration of anyonic statistics within an anyon collider.
\end{abstract}


\maketitle

Out-of-equilibrium (OE) current noise is a valuable tool to explore strongly
correlated mesoscopic conductors and circuits, especially in the high
frequency domain, where it unveils underlying dynamics and models
\cite{blanter_buttiker,FDT_rogovin_scalapino_74,FF_noise_contact_Glattli_PRL_2007,*deblock_06,
lee_levitov,gabelli_violation,trauzettel_group,bena,FF_hall,remark_martin,
zamoum_souquet,FF_noise_DCB_altimiras_PRL_2014,ines_portier_2015,ines_cond_mat_prb,ines_degiovanni_2016}.
It is a major tool in electron quantum optics where it is essential for
characterizing quantum states of electrons
\cite{tomography_Grenier_2011,*tomography_glattli_2014,*tomography_degiovanni_feve_2019}
or of emitted photons 
\cite{hofheinz_photons_11,*squeezing_reulet_PRL_2013,ines_gwendal}. It also
unveils fascinating collective phenomena within strongly correlated
conductors 
as fractional charges
\cite{saminad_ines,ines_cond_mat_prb,christian_photo_2018,ines_gwendal}
and statistics in the fractional quantum Hall effect (FQHE) \cite{ines_prl,*kim_hbt} or charge
splitting in the integer quantum Hall effect (IQHE)
\cite{ines_schulz_tierce,exp_frac,trauzettel_group}. 

The effect of strong correlations in such systems
calls for quantum laws of electronic transport
independent on interactions and the microscopic model of the system.
At equilibrium, 
the fluctuation-dissipation theorem (FDT) which uses the differential
conductance at zero voltage is such a robust
law even in the presence of a nonlinear current at
high voltages \cite{callen_welton_1951}.
%

In the OE regime, fluctuation-dissipation
relations (FDRs) have been long studied 
but mostly at zero frequency \cite{FDT_zero,hekking_subgap_noise_application_perturbative_approach}. A widely used perturbative OE
FDR for high-frequency noise, expressed in terms of the dc curent,
has been derived by Rogovin and Scalapino \cite{FDT_rogovin_scalapino_74}
for independent particles.
Assuming also an initial thermalization, we have extended this FDR to
strongly correlated conductors and 
quantum circuits \cite{bena,zamoum_souquet,ines_cond_mat_prb,ines_degiovanni_2016}, permitting as well a departure from current inversion symmetry to which Ref.\cite{FDT_rogovin_scalapino_74} is restricted.

Finally, in a general OE situation, including a multi-terminal setup with time-dependent voltages,
we have derived universal non-perturbative FDRs 
\cite{ines_philippe_group} which concern only asymmetries between the emission 
(positive frequency) and absorption (negative frequency) parts of the noise spectrum, expressed in terms of OE
non-linear admittance elements.

\begin{figure}[htbp]
\begin{center}
	\includegraphics[width=8cm]{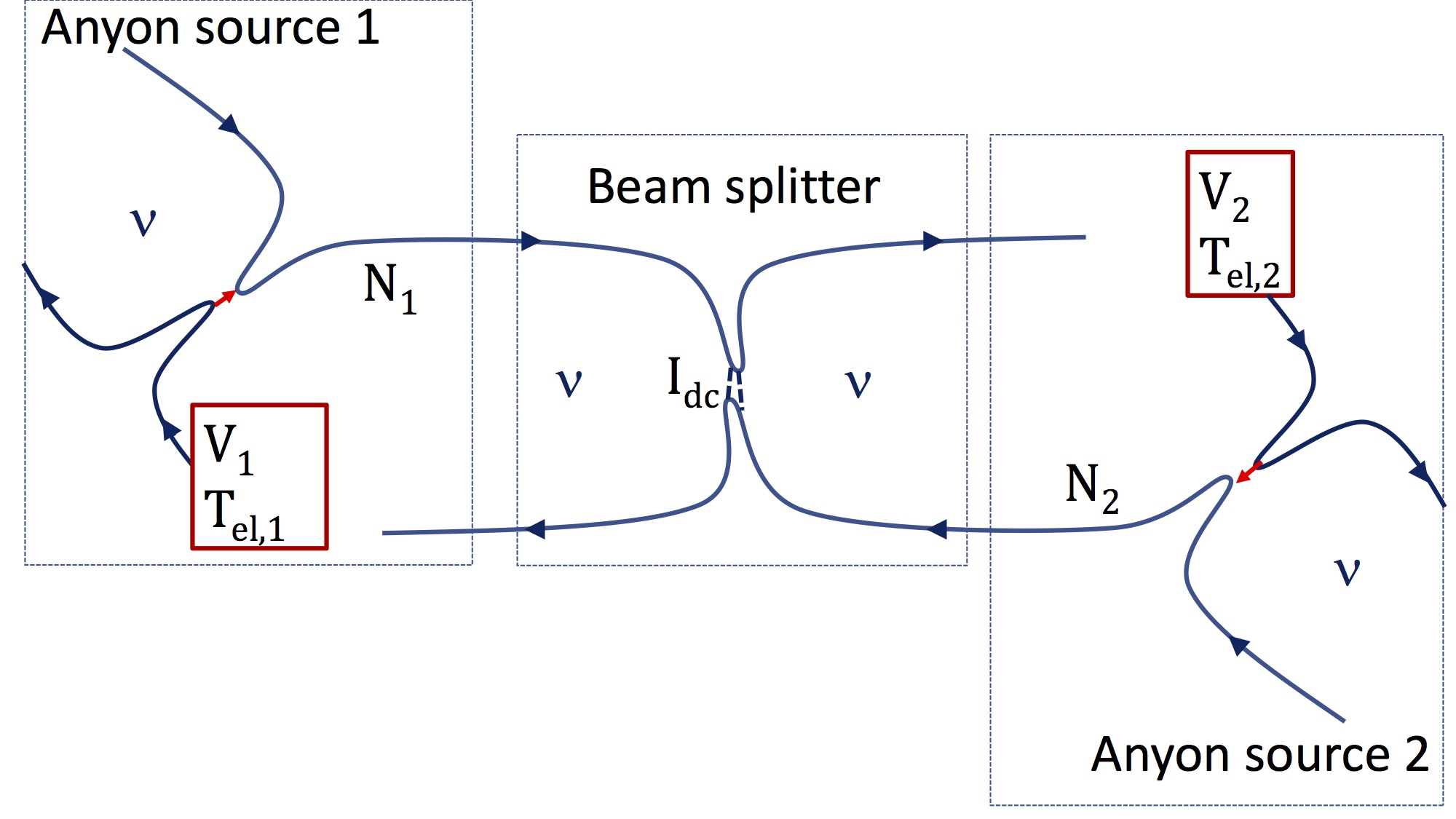}
	\caption{\label{fig:anyon-collider} An anyon collider setup in the
	FQHE. Two QPCs at possibly different
temperatures $T_{\mathrm{el},1}$ and $T_{\mathrm{el},2}$, are subject to 
dc-biases $V_1$ and $V_2$. They inject $N_1$
and $N_2$ anyons into the
upper/down edges which collide at the central QPC. The 
 finite frequency
noise of the backscattering current $I_{dc}$ obeys 
the OE FDRs, independently on the (incompressible) fractional filling factor $\nu$ and the microscopic model.}
\end{center}
\end{figure}

However, the recent developments of interferometry experiments involving
OE stationary sources in the FQHE, such as the anyon
collider shown on Fig. \ref{fig:anyon-collider} 
used to probe the non-trivial statistics of anyons
emitted by non-trivial
sources \cite{fractional_statistics_gwendal_science_2020}, calls for an in-depth
exploration of new FDRs for the full finite frequency noise, 
valid in absence of an
initial thermal state. 
In this paper, we derive perturbative OE FDRs for the full finite
frequency noise without assuming initial thermalization
\cite{non_thermal_TLL_fujisawa_PRB_2016,thermal_group,rev_optics,trauzettel_group,
out_of_equilibrium_bosonisation_eugene_PRL_2009}.
We show that, as long as the perturbative approach remains valid, high
frequency non-symmetrized noise is not fully determined by the dc
current, as in the initially thermalized case
\cite{FDT_rogovin_scalapino_74,ines_cond_mat_prb,ines_degiovanni_2016},
but also by its zero frequency counterpart. This new relation
illustrates the power of the OE perturbative approach, since it can be applied to a
variety of situations independently of any underlying microscopic model.

This is especially relevant for the FQHE: the
OE FDRs derived
for the photo-assisted noise \cite{ines_cond_mat_prb} and for the
high-frequency noise under a dc voltage
\cite{ines_cond_mat_prb,ines_degiovanni_2016} have already provided robust
methods implemented in recent experiments to determine the fractional
charge \cite{christian_photo_2018,ines_gwendal} for filling factors $\nu$
which are not simple fractions, though no experimental signature of the validity of the generic effective
models was observed\cite{wen_review_FQHE_1992,sukho_mac_zhender_PRB_2009}.
We illustrate furthermore the interest of the novel OE FDRs derived here
for the anyon collider. We show that they give 
access to effective interaction-dependent
parameters which are important for the exploration of
anyonic statistics, proposed in Ref. \cite{halperin_HBT_FQHE_2016} and recently implemented in
Ref. \cite{fractional_statistics_gwendal_science_2020}. 
\paragraph{Model} The underlying Hamiltonian of the OE perturbative approach in the 
stationary regime \cite{ines_eugene,ines_cond_mat_prb},
\begin{equation}
	\label{Hamiltonian} 
	\mathcal{H}(t)=\mathcal{H}_0 +e^{-i\omega_Jt} \hat{A}+ e^{i\omega_Jt}\hat{A}^{\dagger} ,
\end{equation} 
involves unspecified Hamiltonian $\mathcal{H}_0$ and 
perturbing operator $\hat{A}$. The
Josephson-like frequency $\omega_J$ must enter only through 
$e^{i\omega_J t}$ in $\mathcal{H}(t)$ 
and is added on top of other dc drives already present
in the system. 
For concreteness, we will focus here on charge transport, though the
theory extends beyond that. We thus assume that there is a charge
operator $\hat{Q}$, conserved by $\mathcal{H}_0$, translated
by a model-dependent 
charge $e^*$ when acting upon by $\hat{A}$. Then, Eq. \eqref{Hamiltonian} implies
that:
\begin{equation}
	\label{eq:current}
	\partial_t \hat{Q}=\;\hat{I}(t)\! =\! \frac{ie^*}{\hbar}\left(
		e^{-i\omega_Jt}\;\hat{A}-e^{i\omega_Jt}\; 
\hat{A}^{\dagger}\right).
\end{equation}
This is true when $\hat{A}$ contains the unitary operator $e^{i\hat{\varphi}}$ 
where the phase operator $\hat{\varphi}$ obeys
$[\hat{\varphi},\hat{Q}]=e^*$. 
In many situations, 
$\partial_t\hat{\varphi}$ obeys a Josephson-type relation with $e^*$
instead of $2e$
\cite{expt_symm,christian_photo_2018,ines_gwendal}:
\begin{equation}
	\label{Jrelation}
\omega_J=\frac{e^*}{\hbar}V_{dc},
\end{equation} 
where $V_{dc}$ is the voltage bias. Quantum averages, denoted by
$\langle ...\rangle$, are taken over a stationary
OE initial
density operator $\rho_0$ ($[\rho_0,\mathcal{H}_0]=0$) thereby
corresponding to 
non-thermal occupation probabilities of many-body
$\mathcal{H}_0$'s eigenstates. These can for example arise from
temperature and dc-voltages biases.

Let us give some
examples. In tunneling
junctions between two similar or different (hybrid) conductors, such as
NIN or SIN junctions \cite{hekking_subgap_noise_application_perturbative_approach}, $\hat{A}$ and $\hat{I}(t)$ respectively correspond
to the tunneling
and electrical current operators. In Josephson junctions, 
$\hat{I}(t)$ is either the quasiparticle ($e^*=e$) or the
pair current ($e^*=2e$). But the form in
Eq. \eqref{Hamiltonian} goes beyond the transfer Hamiltonian approach, as
$\mathcal{H}_0$ is not split into right and left terms, so that it can
incorporate all relevant screened Coulomb interactions.
One can also include in $\mathcal{H}_0$ and $\hat{A}$ strong coupling to a linear or a non-linear
electromagnetic environment.

In the IQHE or the FQHE at
arbitrary incompressible filling factors $\nu$, $\hat{A}$ corresponds to a weak spatially
extended backscattering of electrons or quasiparticles with a fractional charge $e^*$ through a QPC, acting as a beam splitter, and
$\hat{I}(t)$ is the backscattering current. The unperturbed
Hamiltonian $\mathcal{H}_0$ may include
edge reconstruction or
inhomogeneous Coulomb interactions \cite{ines_schulz_tierce}, or even
extended tunneling processes between
counter-propagating edges. As those emanate from different contacts, such processes may not be sufficient to ensure their equilibration
 \cite{wen_review_FQHE_1992}, a situation one could address as well. 

One may also consider OE quasiparticle sources, such as quantum
dots acting as energy filters or biased QPCs. 
As will be illustrated later in the anyon
collider depicted on 
Fig. \ref{fig:anyon-collider}, the Josephson-type relation in
Eq. \eqref{Jrelation} may break down, motivating us to keep $\omega_J$ as a free parameter.

\paragraph{Main OE relations:}
Letting $\delta\hat I_{\mathcal{H}}(t)=\hat I_{\mathcal{H}}(t)-I_{dc}(\omega_{\mathrm{J}})$, 
where the subscript $\mathcal{H}$ refers to the Heisenberg representation with respect 
to $\mathcal{H}(t)$ in Eq. \eqref{Hamiltonian}, we focus on the current noise: 
\begin{equation}
\label{formal_noise}
{S}(\omega_J;t)=\langle \delta\hat I_{\mathcal{H}}(0)\delta\hat
I_{\mathcal{H}}(t)\rangle\,.
\end{equation}
To express $S$ at second-order in $\hat{A}$, we 
replace $\delta\hat I_{\mathcal{H}}(t)$ by $\hat
I_{\mathcal{H}_0}(t)$, or, in Eq. \eqref{eq:current}, 
$\hat{A}_{\mathcal{H}}(t)$ by $\hat{A}_{\mathcal{H}_0}(t)=e^{i\mathcal{H}_0
t} \hat{A}\,e^{-i\mathcal{H}_0 t}$. We obtain these
two building blocks:
\begin{subequations}
\begin{align}
\label{Xup_Xdown:1}
\hbar^2 X_{\rightarrow}(t)&=\langle\hat{A}_{\mathcal{H}_0}^{\dagger}(t)\hat{A}_{\mathcal{H}_0}(0)\rangle\\
\label{Xup_Xdown:2}
\hbar^2 X_{\leftarrow}(t)&=\langle\hat{A}_{\mathcal{H}_0}(0)\hat{A}_{\mathcal{H}_0}^{\dagger}(t)\rangle.
\end{align}
\end{subequations}
Being evaluated in the OE regime characterized by $\mathcal{H}_0$
and $\hat{\rho}_0$, these are OE correlators which don't satisfy any kind of detailed balance 
equations. They
determine the current noise in Eq. \eqref{formal_noise} and its Fourier transform at $\omega$:

\begin{subequations}
\begin{align}
\label{noise_DC_initial:time}
S(\omega_J;t)/e^{*2} &\simeq e^{-i\omega_Jt} X_{\rightarrow}(-t)+e^{i\omega_Jt}X_{\rightarrow}(t)\\
\label{noise_DC_initial:frequency}
S(\omega_J;\omega)/e^{*2} &\simeq
X_{\rightarrow}(\omega_J-\omega)+X_{\leftarrow}(\omega_J+\omega)\,.
\end{align}
\end{subequations}
In particular, the zero frequency noise reads:
\begin{equation}
S(\omega_J;\omega=0)/e^{*2}\simeq X_{\rightarrow}(\omega_J)+X_{\leftarrow}(\omega_J),
	\label{dcnoise}
\end{equation} 
and the dc average current 
\begin{equation}
I_{dc}(\omega_{\mathrm{J}})=\langle \hat{I}_{\mathcal{H}}(t)\rangle
	\simeq
	e^*(X_{\rightarrow}(\omega_{\mathrm{J}})-X_{\leftarrow}(\omega_{\mathrm{J}}))\,
	\label{formal_average}
\end{equation} 
can be interpreted as the difference of two transfer rates
$X_{\rightarrow},X_{\leftarrow}$ in opposite 
directions \cite{ines_cond_mat_prb}.

Then, at a finite frequency $\omega$, the rescaled noise in Eq.
\eqref{noise_DC_initial:frequency} is a sum of these transfer rates evaluated at
two effective potential drops in two opposite directions
$\pm\omega_J-\omega$. 
A transfer of a charge $e^*$ in each direction is associated
with the emission (resp. absorption) of a photon if $\omega>0$ (resp. $\omega<0$)
by the correlated many-body eigenstates,
thus the effective
potential $\pm\omega_J-\omega$ decreases (resp. increases) with respect to
$\pm\omega_J$. 

Comparing Eq. \eqref{noise_DC_initial:frequency} to
Eqs. \eqref{dcnoise},\eqref{formal_average}, we derive the central result
of this paper, an OE FDR expressing the OE current noise at finite frequency
in terms of OE current average and noise at zero frequency \cite{excess}:
\begin{align}
	\label{FDR_initial}
2 S(\omega_J;\omega) &= S(\omega_J+\omega;0)+S(\omega_J-\omega;0)\nonumber\\
&-e^*{I}_{dc}(\omega_J+\omega)+e^*{I}_{dc}(\omega_J-\omega)\,.
\end{align}
Note that the first and second lines on the
r.h.s. yield the symmetric and anti-symmetric parts of the noise
$2S^{\pm}(\omega_J;\omega)=S(\omega_J,\omega)\pm S(\omega_J,-\omega)$.
The high-frequency behavior of $S^+$
is indeed totally determined by its dependence on the dc bias at zero
frequency:
\begin{equation}
2S^+(\omega_J;\omega)=S^+(\omega_J+\omega;0)+S^+(\omega_J-\omega;0)\,.
	\label{noise_DC_initial_sym_0}
\end{equation}
Moreover, using the exact relation  
\cite{ines_philippe_group,bena} 
$S^-(\omega_J;\omega)= -2 \hbar\omega \Re(Y(\omega_J,\omega))$
connecting the anti-symmetric part of the noise to the
OE admittance 
$Y(\omega_J,\omega)$ \cite{admittance_definition}, Eq.
\eqref{FDR_initial} enables us to extend the validity of the relation
\begin{equation}
	\label{admittance}
2 \hbar\omega \Re(Y(\omega_J,\omega))=e^*({I}_{dc}(\omega_J-\omega)-{I}_{dc}(\omega_J+\omega))
\end{equation} 
beyond the hypothesis of initial thermalization 
adopted in
Refs. \cite{ines_eugene,ines_cond_mat_prb,ines_degiovanni_2016}. Since
the Kramers-Kronig relation also yields $\Im(Y(\omega_J;\omega))$ in terms
of ${I}_{dc}$, the admittance $Y(\omega_J,\omega)$ is totally determined 
by the dc-current/voltage characteristic.

The heart of our perturbative approach, underlying the previous the relations, is the fact that OE current and noise can be expressed only through the two OE correlators $X_{\rightarrow}$ and $X_{\leftarrow}$ in Eqs. \eqref{Xup_Xdown:1},\eqref{Xup_Xdown:2}. These are generally independent, as we don't impose any of two hypothesis generically adopted: an odd dc current and thermalization. We can formulate separately these two restrictions, not adopted here, through two links between $X_{\rightarrow}$ and $X_{\leftarrow}$. The first one extends the particle-hole symmetry to strongly correlated systems \cite{ines_cond_mat_prb}:
 \begin{equation}
 \label{inversion}
X_{\rightarrow}(\omega_{\mathrm{J}}) =X_{\leftarrow}(-\omega_J). \end{equation} Thus the transfer rate in one direction is obtained by reversing the sign of the dc drive, so that the dc current in Eq.(\ref{formal_average}) becomes odd: $I_{dc}(\omega_{\mathrm{J}})= -I_{dc}(-\omega_{\mathrm{J}})$ and the noise in Eq.(\ref{noise_DC_initial:frequency}) is even/$\omega_J$: $S(\omega_J;\omega)=S(-\omega_J;\omega)$.  

The second link expresses thermalization at an electronic temperature 
$T_{\mathrm{el}}=1/k_B\beta$ : $X_{\rightarrow}(\omega)=e^{\beta\omega}X_{\leftarrow}(\omega)$. In that case, the OE FDR \eqref{FDR_initial} reduces to
the previously obtained
\cite{ines_cond_mat_prb,ines_degiovanni_2016} FDR:
 \begin{align}
	 \label{noise_thermal}
	 S(\omega_J;\omega)/e^{*2}&=[1+N(\omega_J+\omega)]I_{dc}(\omega_J+\omega)\nonumber\\
	 &+N(\omega_J-\omega)I_{dc}(\omega_J-\omega)
\end{align}
in which $N(\omega)=(e^{\beta\omega}-1)^{-1}$, thereby repositioning a
long stream of model-dependent derivations of this relation
\cite{bena,zamoum_souquet,deblock_06,hofheinz_photons_11,trauzettel_group,bena,FF_hall}
into an unified framework. Note that Rogovin and Scalapino's FDR
\cite{FDT_rogovin_scalapino_74} is recovered from \eqref{noise_thermal}
by considering the symmetric noise: $S^+(\omega_J;\omega)=e^*\sum_{\pm}
\coth\left[\beta(\omega_J\pm\omega)/2\right]I_{dc}(\omega_J\pm\omega)$,
which we have extended beyond its original context and 
without assuming Eq.(\ref{inversion}) \cite{supplemental}.
Indeed, for an initial thermal state, the dc current in
Eq. \eqref{formal_average}, though not odd, has the sign of the dc
bias \cite{ines_cond_mat_prb}: 
\begin{equation}
	\label{sign}
\omega_J I_{dc}(\omega_J)\geq 0,
\end{equation}
but in the general OE case, the current may have the opposite sign of 
$\omega_J$ \cite{hekking_subgap_noise_application_perturbative_approach}.

 Also, two important generic features, obtained at zero and finite frequencies, follow from Eq. \eqref{noise_thermal} at a very low temperature:
the Poissonian statistics and the existence of a threshold for the
emitted noise at $\omega>\omega_J$ \cite{ines_philippe_group}.
We now exploit Eq. \eqref{FDR_initial} to show their common origin and their breakdown for initial OE states. For this, we use properties of $X_{\rightarrow,\leftarrow}(\omega_J)$ in
Eqs. \eqref{Xup_Xdown:1},\eqref{Xup_Xdown:2} derived from their spectral decomposition \cite{ines_cond_mat_prb}. Indeed, $X_{\rightarrow,\leftarrow}(\omega_J)\geq 0$, so that the zero-frequency noise in Eq. \eqref{dcnoise}, compared to Eq. \eqref{formal_average}, obeys:
\begin{equation} 
	\label{super}
S(\omega_J;0) \geq e^* |I_{dc}(\omega_J)|.
\end{equation}
This leads to a lower bound on the high-frequency noise in Eq.
\eqref{FDR_initial} ($\Theta$ is the Heaviside function):
\begin{equation}
\label{FDR_bound}
2 S(\omega_J;\omega) \geq \sum_{\pm}e^*\Theta(\mp
	I_{dc}(\omega_J\pm\omega))|I_{dc}(\omega_J\pm \omega)|\,.
\end{equation}
Let's consider first the case when the system is initially in the ground many-body eigenstate
of $\mathcal{H}_0$. Then, by spectral decomposition, we can show that only one transfer rate
survives ($X_{\rightarrow}(\omega_J<0)=X_{\leftarrow}(\omega_J>0)=0$),
so that Eq. \eqref{sign} holds, and Eqs. \eqref{noise_DC_initial:frequency},\eqref{formal_average} imply that the inequality \eqref{super} 
reduces to an equality: zero-frequency noise is Poissonnian. As a consequence, \eqref{FDR_bound} is also saturated, from which one infers the threshold for the
emission noise at $\omega_J>0$: $S(\omega_J;\omega>\omega_J)=0$. Therefore, single charge transfer processes are Poissonnian and impose energy
conservation underlying the threshold. 

These two features are violated 
when considering OE initial states: the inequality in 
Eq. \eqref{super} is strict, leading to a 
super-Poissonian zero frequency noise. So is the inequality in Eq. \eqref{FDR_bound}, smoothing out the
threshold at $\omega_J$, due to the non-vanishing emission noise above
$\omega_J$: $S(\omega_J;\omega>\omega_J)>0$.
These purely OE effects
persist even at vanishing temperatures. In order
to distinguish them from thermal fluctuations, which also lead to
strict inequalities in Eqs. \eqref{super} and \eqref{FDR_bound}
(see Eqs. \eqref{sign}, \eqref{noise_thermal} with a finite
$T_{\mathrm{el}}$), let us deduce the OE noise at
$\omega_J=0$ from Eq. \eqref{FDR_initial}. For simplicity, we assume that current inversion symmetry, thus Eq.(\ref{inversion}), holds, so that we get:
\begin{equation}
	\label{S_inv}
S(\omega_J=0;\omega)=S(\omega_J=\omega;\omega=0)-e^*I_{dc}(\omega_J=\omega)
\end{equation}
This shows that a
finite emission noise $S(\omega_J=0;\omega>0)$ quantifies
deviations both from the Poissonnian regime and from initial thermalisation, for which it would vanish. 
\paragraph{Applications}
The FRs are alternative laws in the OE regime to the equilibrium FDT, thus provide similarly a robust test of analytical, numerical or experimental results for OE noise. 
One can, inversely, test the validity of the underlying hypotheses of our perturbative approach by checking Eqs. \eqref{FDR_initial} and \eqref{noise_DC_initial_sym_0} \cite{gabelli_violation}, whereas the
signature of a departure from initial thermalization \cite{pierre_equilibration_IQHE_Nature_2010} would be a violation of 
Eq. \eqref{noise_thermal}.
In strongly correlated conductors with OE initial many-body states, a key issue is to 
determine $\omega_J$ in term of the experimentally controlled parameters,
such as dc-voltages and temperatures when $e^*\neq e$ or when
the Josephson-type relation Eq. \eqref{Jrelation} breaks down.

This can be achieved either by measuring the admittance, using Eq. \eqref{admittance}, or by measuring the noise both at
finite and zero frequency, using Eqs. \eqref{FDR_initial},\eqref{noise_DC_initial_sym_0}. One can infer $\omega_{J}$ from the
coincidence of the functions of $\omega$ on both sides of these OE FDRs.

First, these methods could be especially relevant for thermoelectricity \cite{thermal_group}.
The determination of $\omega_J$ provides the
voltage drop across a strongly correlated junction in presence of a temperature
gradient $\Delta T$. In particular, by imposing $I_{dc}(\omega_J)= 0$, 
it offers a method based on current noise measurement to infer
the Seebeck coefficient from $\omega_J/\Delta T$. Note that
at zero bias voltage, the temperature gradient
$\Delta T$ generates a thermoelectric current $I_{dc}(\omega_J=0)\neq
0$ \cite{ines_cond_mat_prb}.

Second, the determination of $\omega_J$ is an especially acute question in the FQHE context, which goes beyond 
that of the fractional charge $e^*$ using Eq. \eqref{Jrelation} when
valid, as in recent experiments
\cite{christian_photo_2018,ines_gwendal}. 
The important point is that,
at a given incompressible
filling factor $\nu$, for example $2/3$, the theoretical description by effective models 
cannot favor one among multiple competing candidates, which may 
even predict different values of $e^*$ \cite{sukho_mac_zhender_PRB_2009}. As of now, because of
Coulomb-induced
non-universal effects such as edge reconstruction, there is no clear
agreement between experiments
\cite{christian_photo_2018,ines_gwendal} and effective models, predicting power laws
\cite{wen_review_FQHE_1992}. 
In this context, the OE FDR can help us sort out, among the various
models, the most suitable one for the experimental data.

Let us illustrate this point in an anyon collider, to show how can the determination of $\omega_J$ help us to pinpoint the
best candidate model.

As depicted on Fig.
\ref{fig:anyon-collider}, two dc-biased QPCs inject anyons with a fractional charge $e^*$, characterized by number operators $\hat{N}_{1,2}$ and averages ${N}_{1,2}$, which collide on the central
QPC. Since equilibrium reservoirs are replaced by OE sources, the backscattering noise obeys the OE FDRs given by
Eqs. \eqref{FDR_initial} and \eqref{noise_DC_initial_sym_0}, but not that given by
Eq. \eqref{noise_thermal} \cite{supplemental}. 
Let's adopt for the edge states, as in Ref.\cite{halperin_HBT_FQHE_2016}, an effective model characterized by two free parameters $\lambda,\delta$ which need to be known to fix the model \cite{sukho_mac_zhender_PRB_2009}. While $\lambda$ refers  to an effective dimensionless charge, $\delta$ monitors the statistical phase of quasiparticles. In case $\nu=1/(2n+1)$, one has $\lambda=\delta=\nu$, but $\lambda,\delta$ may be renormalized by Coulomb interactions and edge reconstruction, whose role can be evaluated by determining experimentally $\lambda,\delta$. Importantly, $\lambda,\delta$, that intervene directly in the cross-correlations of the anyon collider \cite{halperin_HBT_FQHE_2016,fractional_statistics_gwendal_science_2020}, affect their interpretation in terms of anyonic statistics. 

In Ref.\cite{halperin_HBT_FQHE_2016}, $\lambda$ renormalizes $\hat{N}_1,\hat{N}_2$ in the OE part of $\hat{A}$: $\hat{A}\rightarrow
e^{2i\pi\lambda(\hat{N}_1-\hat{N}_2)}\hat{A}$. This derives from the equation of motion method for bosonic fields with boundary conditions fixed by $\hat{N}_1,\hat{N}_2$ \cite{ines_schulz_tierce}, whose higher cumulants are taken into account within the so-called OE bosonisation \cite{out_of_equilibrium_bosonisation_eugene_PRL_2009}. If the QPCs are tuned at weak transmissions and low temperatures, $\hat{N}_{1,2}$ are
Poissonnian, so that their cumulants are proportional to the injected average currents $I_{1,2}=d{N}_{1,2}/dt$, inducing an effective dc drive:
\begin{equation}\label{new}
\omega_J=\frac{2\pi}{e^*}\sin(2\pi\lambda)I_-,
\end{equation} where $I_-=I_1-I_2$. Due to the strongly correlated Hall liquid in the sources, $I_{1},I_{2}$ have a non-linear behavior on $V_1,V_2$, and so does $\omega_J$, which then violates the Josephson-type relation in Eq.(\ref{Jrelation}) (if $V_{dc}=V_1-V_2$).
By using the OE FDR to determine $\omega_J$, and assuming $e^*$ is already inferred from intrinsic noise of the QPCs, one can determine $\sin(2\pi\lambda)$, thus $\lambda$, from Eq.(\ref{new}), as $I_{1},I_{2}$, thus $I_-$, can be measured directly in the outgoing edges \cite{fractional_statistics_gwendal_science_2020}. Indeed, we can show that $\lambda$ describes plasmonic propagation between the
injection point and the central QPC, thus is related it to the dc
conductance by using the scattering approach for plasmons 
\cite{ines_schulz_tierce}. 
One can infer the second parameter $\delta$ from the model-dependent expressions of the dc current and zero-frequency noise in Ref.\cite{halperin_HBT_FQHE_2016}: 
\begin{subequations}
	\label{expressions}
\begin{align}
	\label{expression:a}
	I_{dc}(\omega_J)&= C'\sin( \pi \delta)\, \Im(\omega_++i\omega_J)^{2\delta-1}\\
	\label{expression:b}
S(\omega_J;\omega=0)&= e^*C'\cos(\pi \delta)\, \Re(\omega_++i\omega_J)^{2\delta-1}.
\end{align}
\end{subequations}
 Here $C'$ is a prefactor, $\omega_J$ given by Eq.(\ref{new}), and
$\omega_+=4\pi\sin^2(\pi\lambda)\,I_+/e^*$, with $I_{+}=I_{1}+ I_{2}$.
Though $\delta$ controls the power law, this is not an easy way to extract it, so we propose an alternative way. We notice first that, compared to equilibrium reservoirs, the validity domain of 
perturbation is extended: for high enough $\omega_+$, one can lower $\omega_J$ down to $0$ by injecting equal
currents $I_1=I_2$ through tuning $V_1\simeq V_2$.
This is precisely the regime where anyonic statistics is best revealed
 \cite{halperin_HBT_FQHE_2016}. Then using Eqs. \eqref{expression:a},\eqref{expression:b}, one has :
\begin{equation}
S(\omega_J=0;\omega=0)=e^*
\frac{\cot(\pi \delta)}{1-2\delta}\cot(\pi \lambda)\,
	I_+\left(\frac{\partial I_{dc}}{\partial{I_-}}\right)_{I_-=0},\nonumber
\end{equation}
 proportional to the total
injected current $I_+$ and the derivative of $I_{dc}$ at $\omega_J=0$
(depending on $I_+$). The atypical "Fano factor'' $\cot(\pi \delta)\cot(\pi \lambda)/(1-2\delta)$ then provides $\delta$ once $\lambda$ is determined. 
Now we can express explicitly the high frequency backscattering noise in 
Eq. \eqref{noise_DC_initial_sym_0}, by injecting the dc expressions in Eqs. \eqref{expression:a},\eqref{expression:b}. In particular, at $I_-=0$, as current inversion symmetry now holds, we can use Eq. \eqref{S_inv}
with a fixed $\omega_+$: $S(\omega_J=0;\omega)=-C'\Im(-\omega+i\omega_+)^{2\delta-1}$.

\paragraph{Conclusion} In this Letter, we have derived perturbative OE FRs and FDRs
showing that high-frequency noise
is completely determined by zero-frequency transport. 
Due to OE initial states, zero-frequency noise is
super-Poissonnian, and washes out the threshold for the emitted
spectrum above the dc drive.
 The OE FDRs offer experimental tests of their underlying hypothesis \cite{gabelli_violation}, in particular breakdown of initial thermalization. They provide a noise measurement method 
of the Seebeck coefficient in a strongly correlated junction. 
 In the FQHE, the OE FRs permit to probe the fractional charges without relying on the microscopic
model \cite{ines_cond_mat_prb,christian_photo_2018,ines_gwendal} nor on initial thermal equilibrium. The latter
breaks down in the
anyon collider used to prove 
anyonic statistics \cite{halperin_HBT_FQHE_2016,fractional_statistics_gwendal_science_2020}. The high-frequency backscattering noise does not obey the previously FDRs \cite{ines_cond_mat_prb,ines_degiovanni_2016} but the OE
FDR derived in this paper, which offers
a protocol to extract a non-universal parameter that depends
on the structure of the edge channels and enters anyonic statistics.
This may prove useful in forthcoming investigations of anyonic statistics through finite-frequency correlations.
Future perspectives include using the OE FDRs for shot noise thermometry 
\cite{thermal_group,ines_average}, as well for thermoelectricity in the anyon collider. Beyond current noise, they can be
applied to the voltage noise across a phase-slip Josephson junction
\cite{hekking_various_T_PRB_2013} as well as to the spin current noise in spin Hall insulators \cite{sassetti_spin_hall_LL_PRB_2014,spin_hall_glazman}. 
\paragraph*{Acknowledgments} The author is grateful to B. Dou\c cot for illuminating suggestions, to P. Degiovanni and D. Est\`eve
for many inspiring comments on the manuscript, as well as to G. F\`eve, D. C. Glattli, F. Pierre and I. Taktak for fruitful discussions.

\section*{Supplemental Material}

\subsection{The perturbative approach}
The Hamiltonian within the present approach reads:
 \begin{eqnarray}
	\label{Hamiltonian_s} 
	\mathcal{H}(t)\!\! &= &\!\! \mathcal{H}_0 +\!\! \;e^{-i\omega_Jt} \hat{A}+ e^{i\omega_Jt} \;\hat{A}^{\dagger} ,
\end{eqnarray} 
where we don't specify $\mathcal{H}_0$ nor the weak operator $\hat{A}$, and $\omega_J$ is a dc drive. In the case of a tunneling junction, and contrary to the transfer Hamiltonian description, we don't decouple right and left sides of the junction, which could circumvent difficulties in the construction of the global Hilbert space \cite{tunneling_non_perturbative_PRB_2018}. 
We recall the minimal conditions for the validity of the theory:
\begin{enumerate}
\item $\hat{A}$ is a weak operator which does not depend on $\omega_J$, and with respect to which second-order expansion is valid and yields non-vanishing results. 
\item The initial density matrix $\hat{\rho}_0$ commutes with the unperturbed Hamiltonian $\mathcal{H}_0$: 
\begin{equation}[\hat{\rho}_0,\mathcal{H}_0]=0,\label{commutation}
\end{equation} thus is diagonal with respect to the stationary OE many-body eigenstates of $\mathcal{H}_0$. 
  \item Letting:
\begin{equation} 
	 \hat{A}_{\mathcal{H}_0}(t)\!\!  =\!\! 
	e^{i\mathcal{H}_0 t} \hat{A}\,e^{-i\mathcal{H}_0 t},
	\label{eq:heisenberg} 
	\end{equation} one requires the following cancellation: 
\begin{eqnarray}
	\label{condition_supp} 
\langle \hat{A}_{\mathcal{H}_0}(t)\rangle=  \langle \hat{A}_{\mathcal{H}_0}(t) \hat{A}_{\mathcal{H}_0}(0) \rangle\!\!  &=&\!\!  0,
\end{eqnarray}  

\end{enumerate}
Let us discuss in more details the last condition, Eq.(\ref{condition_supp}). If the diagonal elements of $\rho_0$ are determined only by the energy of $\mathcal{H}_0$'s eigenstates, we have shown that: $I_{dc}(\omega_J=0)=0$ \cite{ines_cond_mat_prb}. For Josephson junctions, supercurrent is negligible with a dissipative environment or a magnetic field. In case one has a temperature gradient, such a condition on $\rho_0$ is violated, so one has  $I_{dc}(\omega_J=0)\neq 0$ even if Eq. \eqref{condition_supp} holds.

In the paper, focussing on charge transport, we have introduced, for clarity, additional though not necessary requirements, which ensure systematically Eq. \eqref{condition_supp}. Let us explain why. First, we have assumed there is a phase operator $\hat{\varphi}$ such that, implicitly, $\hat{A} =e^{i\hat{\varphi}}\bar{A}$, where $\bar{A}$ is an unspecified operator commuting with $\hat{\varphi}$. Second, we assume the unperturbed Hamiltonian $\mathcal{H}_0$ conserves the charge operator $\hat{Q}$ conjugated to $\hat{\varphi}$, thus $\mathcal{H}_0$ does not depend on $\hat{\varphi}$. A charge $e^*$ is introduced through the commutator: $[\hat{\varphi},\hat{Q}]=e^*$. For transfer of multiple values of charges, either one process associated with one value dominates the others, otherwise the noise is a superposition of terms obeying individually the OE FRs we have shown. Generically, the same charge $e^*$ enters into the Josephson-type relation : $\omega_J={e^*}V_{dc}/{\hbar}$. But one could have, depending on coupling to the bias $V_{dc}$ and the OE setup, a different model-dependent charge $q$ such that  $\omega_J=\frac{q}{\hbar}V_{dc}$. This happens for instance in quantum wires with reservoirs, where $e^*$ depends on interactions, but $q=e$ \cite{ines_epj,bena,trauzettel_group}. We have also shown that the Josephson-type relation itself may be violated, as is the case for the anyon collider. So cannot exclude that the different transferred charges might share the same $\omega_J$; in which case the noise still verifies the OE FDRs.n

In a junction, $\hat{Q}$ corresponds to a charge difference operator, which does not change in absence of perturbing $\hat{A}$. One can show that $\langle \hat{A}_{\mathcal{H}_0}(t)\rangle=0$ because $<e^{i\hat{\varphi_{\mathcal{H}_0}(t)}}>=0$, which expresses precisely this conservation. One has also: $$\langle \hat{A}_{\mathcal{H}_0}(t) \hat{A}_{\mathcal{H}_0}(0) \rangle =X_{\varphi}\langle\bar{A}_{\mathcal{H}_0}(t)\bar{A}_{\mathcal{H}_0}(0) \rangle=0,$$
where $X_{\varphi}=<e^{i\hat{\varphi}_{\mathcal{H}_0}(t)}e^{i\hat{\varphi}_{\mathcal{H}_0}(0)}>$. We have used $X_{\varphi}=0$ as $\mathcal{H}_0$ does not depend on $\hat{\varphi}$, which is a trivial case of gauge invariance with respect to $\hat{\varphi}$ (or $U(1)$ symmetry). Indeed $X_{\varphi}$ vanishes because they correspond to overlaps of states with different charges. In a more general context \cite{ines_cond_mat_prb}, it is sufficient, to get Eq. \eqref{condition_supp}, that the action $\mathcal{S}_0$ associated with $\mathcal{H}_0$, obeys, for any real $c$:
\begin{equation} \label{invariance}\mathcal{S}_0 (\hat{\varphi}+c)\!\!  =\!\! \mathcal{S}_0 (\hat{\varphi}).
\end{equation}
This is indeed a sufficient condition replacing $\varphi$ by an alternative phase operator $\hat{\varphi}_1$ on which $\mathcal{H}_0$ depends, not necessarily in a quadratic form. For instance, $\hat{\varphi}_1$ can then be associated with a linear or even non-linear electromagnetic environment, included in $\mathcal{H}_0$. Then the phase $\hat{\varphi}$ above corresponds to intrinsic electronic degrees of freedom of the junction, thus $\hat{A}$ contains $e^{i\hat{\varphi}}e^{i\hat{\varphi}_1}$. If one considers the charge operator $\hat{Q}_1$ conjugate to $\hat{\varphi}_1$, its time derivative contains $\hat{I}(t)$, in addition to $\delta\mathcal{H}_0/\delta\hat{\varphi}_1$.

\subsection{Inversion symmetry or initial thermalization}
In the perturbative approach, in view of Eq. \eqref{condition_supp}, one ends up with
 the two OE correlators in Eq. \eqref{Xup_Xdown}:
\begin{eqnarray}\label{Xup_Xdown_s}
\hbar^2 X_{\rightarrow}(t)&=&\langle\hat{A}_{\mathcal{H}_0}^{\dagger}(t)\hat{A}_{\mathcal{H}_0}(0)\rangle\nonumber\\
\hbar^2 X_{\leftarrow}(t)&=&\langle\hat{A}_{\mathcal{H}_0}(0)\hat{A}_{\mathcal{H}_0}^{\dagger}(t)\rangle,
\end{eqnarray}
evaluated in the OE regime characterized by $\mathcal{H}_0,\hat{\rho}_0$ to which the system stays close trough perturbation theory. They
determine average current and noise:
\begin{eqnarray}
I_{dc}(\omega_{\mathrm{dc}})&\simeq &e^*\left[X_{\rightarrow}(\omega_{\mathrm{dc}})-X_{\leftarrow}(\omega_{\mathrm{dc}})\right].\label{formal_average_s}\\
S(\omega_J;\omega)/e^{*2}\!\! &\simeq &X_{\rightarrow}(\omega_J-\omega)+X_{\leftarrow}(\omega_J+\omega).\label{noise_DC_initial_s}
\end{eqnarray}
The fact that only two independent correlators enter allows us to establish various links, such as the perturbative OE FDR:
\begin{eqnarray}\label{FDR_initial_s}
2 S(\omega_J;\omega) &= &S(\omega_J+\omega;0)+S(\omega_J-\omega;0)\nonumber\\
&&-e^*{I}_{dc}(\omega_J+\omega)+e^*{I}_{dc}(\omega_J-\omega).
\end{eqnarray}
Once symmetrized with respect to $\omega$, we have obtained:
\begin{equation}
2S^+(\omega_J;\omega)=S^+(\omega_J+\omega;0)+S^+(\omega_J-\omega;0)\,.
	\label{noise_DC_initial_sym_0_s}
\end{equation}
Notice that validity of perturbation can be stated by a weak dc differential conductance $G_{dc}(\omega_J)={dI_{dc}(\omega_J)}/{dV_{dc}}$ compared to a model-dependent scale.

We have not required any of two hypothesis simultaneously adopted by almost all works on finite frequency noise: initial thermalization and an odd dc current. This is hidden in the fact that, within the perturbative approach, $X_{\rightarrow}$ and $X_{\leftarrow}$ are two independent OE correlators. Each of these two restrictions, which we will discuss separately, amounts to impose each time a link between $X_{\rightarrow}$ and $X_{\leftarrow}$.

First, particle-hole symmetry is generically the underlying reason for oddness of the current. But we define an alternative symmetry suitable for strongly correlated systems \cite{ines_cond_mat_prb}, by requiring, for the OE correlators in Eq. \eqref{Xup_Xdown}, $X_{\rightarrow}(t)=X_{\leftarrow}(-t)$. Thus their Fourier transforms, evaluated here at the dc drive, are related to a unique function $X$ :
 \begin{equation}
 \label{inversion}
X_{\rightarrow}(\omega_{\mathrm{J}}) =X(\omega_{\mathrm{J}})= X_{\leftarrow}(-\omega_J).
\end{equation} 
This can be described as inversion symmetry, as the transfer rate in one direction is obtained by reversing the sign of the dc drive. With this hypothesis, and with respect to the dc drive $\omega_J$, the dc current in Eq.(\ref{formal_average_s}) is now odd: $I_{dc}(\omega_{\mathrm{J}})= -I_{dc}(-\omega_{\mathrm{J}})$ and the noise in Eq.(\ref{noise_DC_initial_s}) is even: $S(\omega_J;\omega)=S(-\omega_J;\omega)$. \\
In this case, let's specify Eq. \eqref{FDR_initial_s} to $\omega_J=0$. One has still an OE noise given by:
\begin{equation}\label{S_inv_s}S(\omega_J=0;\omega)=S(\omega_J=\omega;\omega=0)-e^*I_{dc}(\omega_J=\omega).
\end{equation}
One gets also the symmetrized noise with respect to frequency, $S^+(\omega_J=0;\omega)=S^+(\omega_J=\omega;0)$, thus frequency and dc drive exchange their roles; thus one can infer one function from the other, depending on which one is the most easily accessible theoretically or experimentally. \\

 Second, we consider now a the link between the OE correlators in Eq.(\ref{Xup_Xdown_s}) :
\begin{equation}\label{balance}
X_{\rightarrow}(\omega)=e^{\beta\omega}X_{\leftarrow}(\omega).
\end{equation}
Though $\beta$ could acquire a different meaning, this link arises from the choice of an initial thermalization at $T_{el}=1/\beta$:
\begin{equation}\label{app_thermal_supp}
\hat{\rho}_0=\frac{e^{-\beta\mathcal{H}_0}}{Tr \; e^{-\beta\mathcal{H}_0}}.
\end{equation}
This leads us to recover the OE-FDR for the FF non-symmetrized \cite{ines_cond_mat_prb,ines_degiovanni_2016} noise in terms of the dc average current:
\begin{eqnarray}\label{noise_thermal_s}
S(\omega_J;\omega)/e^*&=&[1+N(\omega_J+\omega)]I_{dc}(\omega_J+\omega)\nonumber\\&&+N(\omega_J-\omega)I_{dc}(\omega_J-\omega),
\end{eqnarray}
where $N(\omega)=(e^{\beta\omega}-1)^{-1}$. We have also obtained in the FQHE or in a conductor connected to an electromagnetic environment  \cite{bena,zamoum_souquet}.
 Here we choose to give a a different derivation of Eq. \eqref{noise_thermal}, compared to \cite{ines_cond_mat_prb,ines_degiovanni_2016}, by exploiting directly the novel FDRs. 
Using Eqs. (\ref{formal_average},\ref{balance}), the zero-frequency noise obeys: 
\begin{eqnarray}\label{noise_DC_zero_s}
S(\omega_J;0)/e^* &= &\coth\left(\frac{\beta\omega_J}{2}\right) I_{dc}(\omega_J).
\end{eqnarray}
Injecting it into the expression of the non-symmetrized finite-frequency noise in Eq. \eqref{FDR_initial_s}, we recover Eq. \eqref{noise_thermal_s}. 
One can also get directly the symmetrised noise, injecting Eq. \eqref{noise_DC_zero_s} in Eq. \eqref{noise_DC_initial_sym_0_s}: $S^+(\omega_J;\omega)=e^*\sum_{\pm}
\coth\left[\beta(\omega_J\pm\omega)/2\right]I_{dc}(\omega_J\pm\omega)$. This is the sam form as Rogovin and Scalapino's FDR, here extended to a much larger large family of strongly correlated systems and circuits described by Eq.(\ref{Hamiltonian_s}). It holds beyond the particle-hole symmetry, thus oddness of the current, on which Rogovin and Scalapino have insisted \cite{FDT_rogovin_scalapino_74}. \\
It is only when we assume inversion symmetry, thus Eq. \eqref{inversion}, that, using Eq. \eqref{balance}, we recover the detailed balance equation for a unique transfer rate $X(\omega)=X_{\rightarrow}(\omega)=X_{\leftarrow}(-\omega_J)=e^{\beta \omega}X(-\omega)$, as in Ref.\cite{FDT_rogovin_scalapino_74}. Interestingly, even without inversion symmetry, the equilibrium noise, now given by $S_{eq}(\omega)=S(\omega_J=0;\omega)$, obeys: $S_{eq}(-\omega)=e^{\beta\omega}S_{eq}(\omega)$, as one can show by using directly Eqs. (\ref{FDR_initial_s},\ref{balance}).

\subsection{The anyon collider}

\begin{figure}[htbp]
\begin{center}
	\includegraphics[width=8cm]{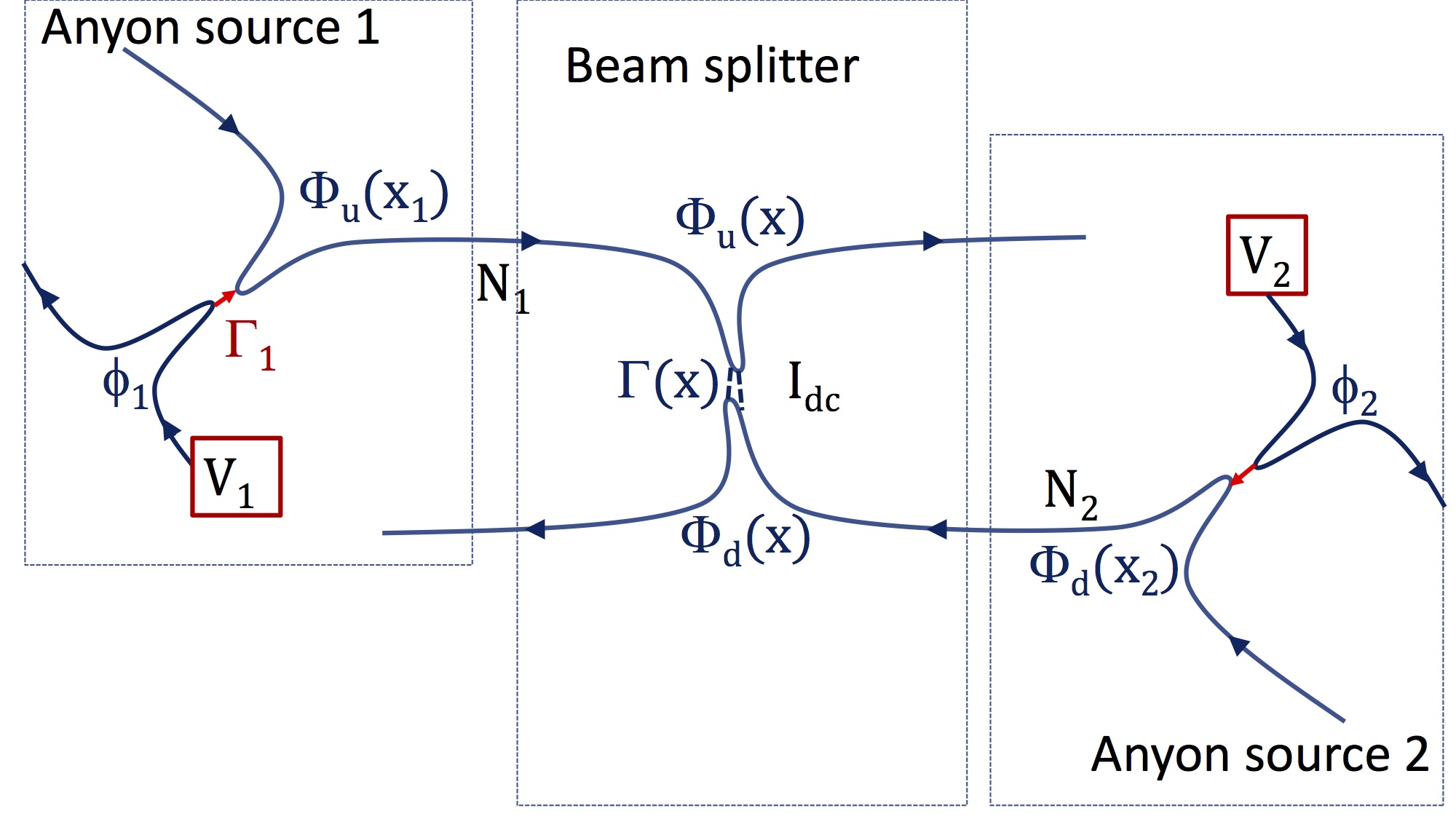}
	\caption{\label{fig:anyon-collider-supp} An anyon collider setup in the
	FQHE. Here we specify, as in \cite{halperin_HBT_FQHE_2016}, to two QPCs with weak backscattering amplitudes $\Gamma_{1,2}$, to $\nu$ a simple fraction with an effective bosonic model, thus four chiral fields intervene here. The injected anyons into the
upper/down edges collide at the beam splitter, where we allow for extended and weak backscattering amplitudes $\Gamma(x)$. The 
 finite frequency
noise of the backscattering current $I_{dc}$ obeys 
the OE FDRs independently on the fractional filling factor $\nu$ and the microscopic model.}
\end{center}
\end{figure}
Let's now discuss the anyon collider in the FQHE addressed in Ref. \cite{halperin_HBT_FQHE_2016} and implemented in Ref. \cite{fractional_statistics_gwendal_science_2020} (see  Fig.(\ref{fig:anyon-collider-supp})). Here we assume simple filling factors $\nu=1/(2n+1)$ for simplicity, and that the three QPCs have weak transmissions. One can associate two commuting bosonic phase fields to the upper and lower edges, $\phi_{u,d}(x)$. Also we assume the injecting QPCs are local in space, so that one introduces two bosonic fields $\phi_{1,2}$, which are evaluated only at the tunneling positions, designated by $x_1$ ($x_2$) for the coordinates on the down (upper) edge. We let $\Gamma_1,\Gamma_2$ the corresponding tunneling amplitudes. In addition to quadratic chiral Hamiltonians corresponding to the four chiral fields $\phi_{u,d},\phi_{1,2}$, which all obey the invariance in Eq.(\ref{invariance}), one describes the central QPC (the beam splitter), by extended backscattering processes with amplitude $\Gamma(x)$ (up to some prefactors):
\begin{equation}\label{A_QPC}
\hat{A}\!\! =\!\!  \int d{{x}}\Gamma(x)e^{i\sqrt{\nu}({\phi}_u(x)-\phi_{d}(x))}.
\end{equation}
One adds, for the two injecting QPCs (up to prefactors):
\begin{eqnarray}\label{injecting_QPC}
\mathcal{H}_{1}&=&\Gamma_{1}e^{i\sqrt{\nu}({\phi}_1-\phi_{d}(x_1))}+h. c.\nonumber\\
\mathcal{H}_{2}&=&\Gamma_{2}e^{i\sqrt{\nu}({\phi}_2-\phi_{u}(x_2))}+h.c..
\end{eqnarray}
One has also to add the linear coupling terms between $\partial_x {\phi}_{1,2}$ to the dc voltages $V_{1,2}$. 
One could carry on perturbation with respect to weak $\Gamma_{1},\Gamma_{2},\Gamma$, but it turns out that one gets divergent results at zero temperature. This divergence was noticed in a similar geometry with a unique QPC, in Ref.\cite{kane_fisher_dilute}, and we have explained its underlying mechanism in Ref.\cite{ines_cond_09}. 

The OE bosonisation approach \cite{out_of_equilibrium_bosonisation_eugene_PRL_2009} has the advantage to take into account the QPCs in a non-perturbative way. It extends the equation of motion method with boundary conditions, we have initiated in \cite{ines_epj,note_OE}.  \\
Boundary conditions are now given by the injected number operators $\hat{N}_{1,2}$. By solving the chiral equations of motion for $\phi{u,d}$, one gets their translation $
\phi_{u,d}+\nu(\hat{N}_1-\hat{N}_2)$, thus $\hat{A}\rightarrow e^{ 2\pi\lambda(\hat{N}_1-\hat{N}_2)}\hat{A}$. The parameter $\lambda$ describes plasmonic propagation along the upper edge, and we can relate it to the dc conductance without the QPCs, using the plasmon approach \cite{ines_epj,note_OE}. Thus $\lambda=1/m$, which is the value of the quantized dc conductance, nonetheless $\lambda$ deviates from this value by edge reconstruction. 

The backscattering current and noise associated with Eq.(\ref{A_QPC}), to second order with respect to the backscattering amplitude $\Gamma(x)$, obey Eq. \eqref{FDR_initial}, with the dc drive $\omega_J={2\pi}\sin(2\pi\lambda)(I_1-I_2)/{e^*}$.
Using the explicit expressions for the dc backscattering current and noise, one can deduce the finite-frequency non-symmetrized noise. In particular, at $I_1=I_2$, thus at $\omega_J=0$, $S(\omega_J=0;\omega)=-C'\Im(-\omega+i\omega_+)^{2\delta-1}$, where $\omega_+=2\pi\sin^2(\pi\lambda)(I_{1}+ I_{2})/e^*$.
If $\omega$ is high enough, we can let $\omega_+=0$ to get the equilibrium noise: $S_{eq}(\omega)=C'\sin(2\pi\delta)\omega^{2\delta-1}$. We notice that  $S(\omega_J=0;\omega)=S_{eq}(\omega)$ whenever $\delta=1$, thus for a linear dc current, or when $\lambda\ll 1$, so that one is close to a thermal state \cite{halperin_HBT_FQHE_2016}, which we can understand through the reduction of the OE contribution $\lambda(\hat{N}_1-\hat{N}_2)$. \\


\begin{thebibliography}{47}%
\makeatletter
\providecommand \@ifxundefined [1]{%
 \@ifx{#1\undefined}
}%
\providecommand \@ifnum [1]{%
 \ifnum #1\expandafter \@firstoftwo
 \else \expandafter \@secondoftwo
 \fi
}%
\providecommand \@ifx [1]{%
 \ifx #1\expandafter \@firstoftwo
 \else \expandafter \@secondoftwo
 \fi
}%
\providecommand \natexlab [1]{#1}%
\providecommand \enquote  [1]{``#1''}%
\providecommand \bibnamefont  [1]{#1}%
\providecommand \bibfnamefont [1]{#1}%
\providecommand \citenamefont [1]{#1}%
\providecommand \href@noop [0]{\@secondoftwo}%
\providecommand \href [0]{\begingroup \@sanitize@url \@href}%
\providecommand \@href[1]{\@@startlink{#1}\@@href}%
\providecommand \@@href[1]{\endgroup#1\@@endlink}%
\providecommand \@sanitize@url [0]{\catcode `\\12\catcode `\$12\catcode
  `\&12\catcode `\#12\catcode `\^12\catcode `\_12\catcode `\%12\relax}%
\providecommand \@@startlink[1]{}%
\providecommand \@@endlink[0]{}%
\providecommand \url  [0]{\begingroup\@sanitize@url \@url }%
\providecommand \@url [1]{\endgroup\@href {#1}{\urlprefix }}%
\providecommand \urlprefix  [0]{URL }%
\providecommand \Eprint [0]{\href }%
\providecommand \doibase [0]{https://doi.org/}%
\providecommand \selectlanguage [0]{\@gobble}%
\providecommand \bibinfo  [0]{\@secondoftwo}%
\providecommand \bibfield  [0]{\@secondoftwo}%
\providecommand \translation [1]{[#1]}%
\providecommand \BibitemOpen [0]{}%
\providecommand \bibitemStop [0]{}%
\providecommand \bibitemNoStop [0]{.\EOS\space}%
\providecommand \EOS [0]{\spacefactor3000\relax}%
\providecommand \BibitemShut  [1]{\csname bibitem#1\endcsname}%
\let\auto@bib@innerbib\@empty
\bibitem [{\citenamefont {Blanter}\ and\ \citenamefont
  {B\"{u}ttiker}(2000)}]{blanter_buttiker}%
  \BibitemOpen
  \bibfield  {author} {\bibinfo {author} {\bibfnamefont {Y.~M.}\ \bibnamefont
  {Blanter}}\ and\ \bibinfo {author} {\bibfnamefont {M.}~\bibnamefont
  {B\"{u}ttiker}},\ } {\bibfield  {journal} {\bibinfo
  {journal} {Phys. Rep.}\ }\textbf {\bibinfo {volume} {336}},\ \bibinfo {pages}
  {1} (\bibinfo {year} {2000})}\BibitemShut {NoStop}%
\bibitem [{\citenamefont {Rogovin}\ and\ \citenamefont
  {Scalapino}(1974)}]{FDT_rogovin_scalapino_74}%
  \BibitemOpen
  \bibfield  {author} {\bibinfo {author} {\bibfnamefont {D.}~\bibnamefont
  {Rogovin}}\ and\ \bibinfo {author} {\bibfnamefont {D.}~\bibnamefont
  {Scalapino}},\ } {\bibfield
  {journal} {\bibinfo  {journal} {Annals of Physics}\ }\textbf {\bibinfo
  {volume} {86}},\ \bibinfo {pages} {1 } (\bibinfo {year} {1974})}\BibitemShut
  {NoStop}%
\bibitem [{\citenamefont {Zakka-Bajjani}\ \emph {et~al.}(2007)\citenamefont
  {Zakka-Bajjani}, \citenamefont {S\'egala}, \citenamefont {Portier},
  \citenamefont {Roche}, \citenamefont {Glattli}, \citenamefont {Cavanna},\
  and\ \citenamefont {Jin}}]{FF_noise_contact_Glattli_PRL_2007}%
  \BibitemOpen
  \bibfield  {author} {\bibinfo {author} {\bibfnamefont {E.}~\bibnamefont
  {Zakka-Bajjani}}, \bibinfo {author} {\bibfnamefont {J.}~\bibnamefont
  {S\'egala}}, \bibinfo {author} {\bibfnamefont {F.}~\bibnamefont {Portier}},
  \bibinfo {author} {\bibfnamefont {P.}~\bibnamefont {Roche}}, \bibinfo
  {author} {\bibfnamefont {D.~C.}\ \bibnamefont {Glattli}}, \bibinfo {author}
  {\bibfnamefont {A.}~\bibnamefont {Cavanna}},\ and\ \bibinfo {author}
  {\bibfnamefont {Y.}~\bibnamefont {Jin}},\ } {\bibfield  {journal}
  {\bibinfo  {journal} {Phys. Rev. Lett.}\ }\textbf {\bibinfo {volume} {99}},\
  \bibinfo {pages} {236803} (\bibinfo {year} {2007})}\BibitemShut {NoStop}%
\bibitem [{\citenamefont {P.~M.~Billangeon}\ and\ \citenamefont
  {Deblock}(2006)}]{deblock_06}%
  \BibitemOpen
  \bibfield  {author} {\bibinfo {author} {\bibfnamefont {H.~B.}\ \bibnamefont
  {P.~M.~Billangeon}, \bibfnamefont {F.~Pierre}}\ and\ \bibinfo {author}
  {\bibfnamefont {R.}~\bibnamefont {Deblock}},\ } {\bibfield  {journal} {\bibinfo
  {journal} {Phys. Rev. Lett.}\ }\textbf {\bibinfo {volume} {96}},\ \bibinfo
  {pages} {136804} (\bibinfo {year} {2006})}\BibitemShut {NoStop}%
\bibitem [{lee()}]{lee_levitov}%
  \BibitemOpen
  \href@noop {} {}\bibinfo {note} {H. Lee and L. S. Levitov, Phys. Rev. B 53,
  7383, 1996}\BibitemShut {NoStop}%
\bibitem [{gab()}]{gabelli_violation}%
  \BibitemOpen
  \href@noop {} {}\bibinfo {note} {P. F\'evrier and J. Gabelli, Nature Comm.{
  \bf 9}, 4940 (2018).}\BibitemShut {Stop}%
\bibitem [{tra()}]{trauzettel_group}%
  \BibitemOpen
  \href@noop {} {}\bibinfo {note} {{B. Trauzettel}, {I. Safi}, {F. Dolcini},
  and {H. Grabert}, Phys. Rev. Lett. {\bf 92}, 226405 (2004). E. Berg, Y. Oreg,
  E.-A. Kim, and F. von Oppen, Phys. Rev. Lett. {\bf 102}, 236402 (2009). I.
  Neder, Phys. Rev. Lett. {\bf 108},186404 (2012).}\BibitemShut {Stop}%
\bibitem [{ben()}]{bena}%
  \BibitemOpen
  \href@noop {} {}\bibinfo {note} {C. Bena and I. Safi, Phys. Rev. B {\bf 76},
  125317 (2007). I. Safi, C. Bena, and A. Cr\'epieux, Phys. Rev. B {\bf 78},
  205422 (2008).}\BibitemShut {Stop}%
\bibitem [{FF_()}]{FF_hall}%
  \BibitemOpen
  \href@noop {} {}\bibinfo {note} {C. de C. Chamon, D. E. Freed, and X. G. Wen,
  Phys. Rev. B 51, 2363 (1995-II) (symmetrized noise). A more general framework
  for non-symmetrized noise was developed in Refs.\cite{bena,trauzettel_group},
  and exploited in: D. Ferraro {\it al}, Phys. Scr. {\bf T151}, 014025 (2012).
  \bibinfo{author}{\bibfnamefont{A.}~\bibnamefont{Braggio}}, {\it et al},
  \bibinfo{journal}{J. Stat. Mech.: Theory Exp}
  \textbf{\bibinfo{volume}{2016}}, \bibinfo{pages}{054010}
  (\bibinfo{year}{2016}).
  \bibinfo{author}{\bibfnamefont{P.}~\bibnamefont{Recher}} {\it al},
  \bibinfo{journal}{Phys. Rev. B} \textbf{\bibinfo{volume}{74}},
  \bibinfo{eid}{235438} ~\bibinfo{numpages}{20},
  (\bibinfo{year}{2006}).}\BibitemShut {Stop}%
\bibitem [{rem()}]{remark_martin}%
  \BibitemOpen
  \href@noop {}{}\bibinfo {note} {For the STM's current noise in A. Lebedev
  {\it et al}, Phys. Rev. B {\bf 71}, 075416 (2005), perturbation is
  invalidated for symmetric voltages on the 1-D wire, as shown in I. Safi,
  arxiv:0908.4382009}\BibitemShut {NoStop}%
\bibitem [{zam()}]{zamoum_souquet}%
  \BibitemOpen
  \href@noop {} {}\bibinfo {note} {R. Zamoum, A. Cr\'epieux, and I. Safi, Phys.
  Rev. B {\bf 85}, 125421 (2012). J.-R. Souquet, I. Safi, and P. Simon, Phys.
  Rev. B {\bf 88}, 205419 (2013).}\BibitemShut {Stop}%
\bibitem [{\citenamefont {Altimiras}\ \emph {et~al.}(2014)\citenamefont
  {Altimiras}, \citenamefont {Parlavecchio}, \citenamefont {Joyez},
  \citenamefont {Vion}, \citenamefont {Roche}, \citenamefont {Esteve},\ and\
  \citenamefont {Portier}}]{FF_noise_DCB_altimiras_PRL_2014}%
  \BibitemOpen
  \bibfield  {author} {\bibinfo {author} {\bibfnamefont {C.}~\bibnamefont
  {Altimiras}}, \bibinfo {author} {\bibfnamefont {O.}~\bibnamefont
  {Parlavecchio}}, \bibinfo {author} {\bibfnamefont {P.}~\bibnamefont {Joyez}},
  \bibinfo {author} {\bibfnamefont {D.}~\bibnamefont {Vion}}, \bibinfo {author}
  {\bibfnamefont {P.}~\bibnamefont {Roche}}, \bibinfo {author} {\bibfnamefont
  {D.}~\bibnamefont {Esteve}},\ and\ \bibinfo {author} {\bibfnamefont
  {F.}~\bibnamefont {Portier}},\ } {\bibfield  {journal}
  {\bibinfo  {journal} {{Physical Review Letters}}\ }\textbf {\bibinfo {volume}
  {112}},\ \bibinfo {pages} {236803} (\bibinfo {year} {2014})}\BibitemShut
  {NoStop}%
\bibitem [{\citenamefont {Parlavecchio}\ \emph {et~al.}(2015)\citenamefont
  {Parlavecchio}, \citenamefont {Altimiras}, \citenamefont {Souquet},
  \citenamefont {Simon}, \citenamefont {Safi}, \citenamefont {Joyez},
  \citenamefont {Vion}, \citenamefont {Roche}, \citenamefont {Est\`eve},\ and\
  \citenamefont {Portier}}]{ines_portier_2015}%
  \BibitemOpen
  \bibfield  {author} {\bibinfo {author} {\bibfnamefont {O.}~\bibnamefont
  {Parlavecchio}}, \bibinfo {author} {\bibfnamefont {C.}~\bibnamefont
  {Altimiras}}, \bibinfo {author} {\bibfnamefont {J.-R.}\ \bibnamefont
  {Souquet}}, \bibinfo {author} {\bibfnamefont {P.}~\bibnamefont {Simon}},
  \bibinfo {author} {\bibfnamefont {I.}~\bibnamefont {Safi}}, \bibinfo {author}
  {\bibfnamefont {P.}~\bibnamefont {Joyez}}, \bibinfo {author} {\bibfnamefont
  {D.}~\bibnamefont {Vion}}, \bibinfo {author} {\bibfnamefont {P.}~\bibnamefont
  {Roche}}, \bibinfo {author} {\bibfnamefont {D.}~\bibnamefont {Est\`eve}},\ and\
  \bibinfo {author} {\bibfnamefont {F.}~\bibnamefont {Portier}},\ } {\bibfield  {journal}
  {\bibinfo  {journal} {Phys. Rev. Lett.}\ }\textbf {\bibinfo {volume} {114}},\
  \bibinfo {pages} {126801} (\bibinfo {year} {2015})}\BibitemShut {NoStop}%
  \bibitem [{ine({\natexlab{a}})}]{ines_cond_mat_prb}%
  \BibitemOpen
 \ \bibinfo {note} {I. Safi, arxiv:1401.5950
  (2014); Phys. Rev. B {\bf{99}}, 045101 (2019).}\BibitemShut {Stop}%
\bibitem [{\citenamefont {Roussel}\ \emph {et~al.}(2016)\citenamefont
  {Roussel}, \citenamefont {Degiovanni},\ and\ \citenamefont
  {Safi}}]{ines_degiovanni_2016}%
  \BibitemOpen
  \bibfield  {author} {\bibinfo {author} {\bibfnamefont {B.}~\bibnamefont
  {Roussel}}, \bibinfo {author} {\bibfnamefont {P.}~\bibnamefont
  {Degiovanni}},\ and\ \bibinfo {author} {\bibfnamefont {I.}~\bibnamefont
  {Safi}},\ } {\bibfield
  {journal} {\bibinfo  {journal} {Phys. Rev. B}\ }\textbf {\bibinfo {volume}
  {93}},\ \bibinfo {pages} {045102} (\bibinfo {year} {2016})}\BibitemShut
  {NoStop}%
\bibitem [{\citenamefont {Grenier}\ \emph {et~al.}(2011)\citenamefont
  {Grenier}, \citenamefont {Herv{\'{e}}}, \citenamefont {Bocquillon},
  \citenamefont {Parmentier}, \citenamefont {Pla{\c{c}}ais}, \citenamefont
  {Berroir}, \citenamefont {F{\`{e}}ve},\ and\ \citenamefont
  {Degiovanni}}]{tomography_Grenier_2011}%
  \BibitemOpen
  \bibfield  {author} {\bibinfo {author} {\bibfnamefont {C.}~\bibnamefont
  {Grenier}}, \bibinfo {author} {\bibfnamefont {R.}~\bibnamefont
  {Herv{\'{e}}}}, \bibinfo {author} {\bibfnamefont {E.}~\bibnamefont
  {Bocquillon}}, \bibinfo {author} {\bibfnamefont {F.~D.}\ \bibnamefont
  {Parmentier}}, \bibinfo {author} {\bibfnamefont {B.}~\bibnamefont
  {Pla{\c{c}}ais}}, \bibinfo {author} {\bibfnamefont {J.~M.}\ \bibnamefont
  {Berroir}}, \bibinfo {author} {\bibfnamefont {G.}~\bibnamefont
  {F{\`{e}}ve}},\ and\ \bibinfo {author} {\bibfnamefont {P.}~\bibnamefont
  {Degiovanni}},\ } {\bibfield  {journal}
  {\bibinfo  {journal} {New J. Phys.}\ }\textbf {\bibinfo {volume} {13}},\
  \bibinfo {pages} {093007} (\bibinfo {year} {2011})}\BibitemShut {NoStop}%
\bibitem [{\citenamefont {Jullien}\ \emph {et~al.}(2014)\citenamefont
  {Jullien}, \citenamefont {Roulleau}, \citenamefont {Roche}, \citenamefont
  {Cavanna}, \citenamefont {Jin},\ and\ \citenamefont
  {Glattli}}]{tomography_glattli_2014}%
  \BibitemOpen
  \bibfield  {author} {\bibinfo {author} {\bibfnamefont {T.}~\bibnamefont
  {Jullien}}, \bibinfo {author} {\bibfnamefont {P.}~\bibnamefont {Roulleau}},
  \bibinfo {author} {\bibfnamefont {B.}~\bibnamefont {Roche}}, \bibinfo
  {author} {\bibfnamefont {A.}~\bibnamefont {Cavanna}}, \bibinfo {author}
  {\bibfnamefont {Y.}~\bibnamefont {Jin}},\ and\ \bibinfo {author}
  {\bibfnamefont {D.~C.}\ \bibnamefont {Glattli}},\ },\href@noop {}
  {\bibfield  {journal} {\bibinfo  {journal} {Nature}\ }\textbf {\bibinfo
  {volume} {514}},\ \bibinfo {pages} {603} (\bibinfo {year}
  {2014})}\BibitemShut {NoStop}%
\bibitem [{\citenamefont {Bisognin}\ \emph {et~al.}(2019)\citenamefont
  {Bisognin}, \citenamefont {Marguerite}, \citenamefont {Roussel},
  \citenamefont {Kumar}, \citenamefont {Cabart}, \citenamefont {Chapdelaine},
  \citenamefont {Mohammad-Djafari}, \citenamefont {Berroir}, \citenamefont
  {Bocquillon}, \citenamefont {Plaçais}, \citenamefont {Cavanna},
  \citenamefont {Gennser}, \citenamefont {Jin}, \citenamefont {Degiovanni},\
  and\ \citenamefont {F\`eve}}]{tomography_degiovanni_feve_2019}%
  \BibitemOpen
  \bibfield  {author} {\bibinfo {author} {\bibfnamefont {R.}~\bibnamefont
  {Bisognin}}, \bibinfo {author} {\bibfnamefont {A.}~\bibnamefont
  {Marguerite}}, \bibinfo {author} {\bibfnamefont {B.}~\bibnamefont {Roussel}},
  \bibinfo {author} {\bibfnamefont {M.}~\bibnamefont {Kumar}}, \bibinfo
  {author} {\bibfnamefont {C.}~\bibnamefont {Cabart}}, \bibinfo {author}
  {\bibfnamefont {C.}~\bibnamefont {Chapdelaine}}, \bibinfo {author}
  {\bibfnamefont {A.}~\bibnamefont {Mohammad-Djafari}}, \bibinfo {author}
  {\bibfnamefont {J.~M.}\ \bibnamefont {Berroir}}, \bibinfo {author}
  {\bibfnamefont {E.}~\bibnamefont {Bocquillon}}, \bibinfo {author}
  {\bibfnamefont {B.}~\bibnamefont {Plaçais}}, \bibinfo {author}
  {\bibfnamefont {A.}~\bibnamefont {Cavanna}}, \bibinfo {author} {\bibfnamefont
  {U.}~\bibnamefont {Gennser}}, \bibinfo {author} {\bibfnamefont
  {Y.}~\bibnamefont {Jin}}, \bibinfo {author} {\bibfnamefont {P.}~\bibnamefont
  {Degiovanni}},\ and\ \bibinfo {author} {\bibfnamefont {G.}~\bibnamefont
  {F\`eve}},\ } {\bibfield  {journal} {\bibinfo
  {journal} {Nat. Commun.}\ }\textbf {\bibinfo {volume} {10}},\ \bibinfo
  {pages} {3379} (\bibinfo {year} {2019})}\BibitemShut {NoStop}%
\bibitem [{\citenamefont {Hofheinz}\ \emph {et~al.}(2011)\citenamefont
  {Hofheinz}, \citenamefont {Portier}, \citenamefont {Baudouin}, \citenamefont
  {Joyez}, \citenamefont {Vion}, \citenamefont {Bertet}, \citenamefont
  {Roche},\ and\ \citenamefont {Esteve}}]{hofheinz_photons_11}%
  \BibitemOpen
  \bibfield  {author} {\bibinfo {author} {\bibfnamefont {M.}~\bibnamefont
  {Hofheinz}}, \bibinfo {author} {\bibfnamefont {F.}~\bibnamefont {Portier}},
  \bibinfo {author} {\bibfnamefont {Q.}~\bibnamefont {Baudouin}}, \bibinfo
  {author} {\bibfnamefont {P.}~\bibnamefont {Joyez}}, \bibinfo {author}
  {\bibfnamefont {D.}~\bibnamefont {Vion}}, \bibinfo {author} {\bibfnamefont
  {P.}~\bibnamefont {Bertet}}, \bibinfo {author} {\bibfnamefont
  {P.}~\bibnamefont {Roche}},\ and\ \bibinfo {author} {\bibfnamefont
  {D.}~\bibnamefont {Esteve}},\ } {\bibfield  {journal}
  {\bibinfo  {journal} {Phys. Rev. Lett.}\ }\textbf {\bibinfo {volume} {106}},\
  \bibinfo {pages} {217005} (\bibinfo {year} {2011})}\BibitemShut {NoStop}%
\bibitem [{\citenamefont {Gasse}\ \emph {et~al.}(2013)\citenamefont {Gasse},
  \citenamefont {Lupien},\ and\ \citenamefont
  {Reulet}}]{squeezing_reulet_PRL_2013}%
  \BibitemOpen
  \bibfield  {author} {\bibinfo {author} {\bibfnamefont {G.}~\bibnamefont
  {Gasse}}, \bibinfo {author} {\bibfnamefont {C.}~\bibnamefont {Lupien}},\ and\
  \bibinfo {author} {\bibfnamefont {B.}~\bibnamefont {Reulet}},\ } {\bibfield  {journal}
  {\bibinfo  {journal} {Phys. Rev. Lett.}\ }\textbf {\bibinfo {volume} {111}},\
  \bibinfo {pages} {136601} (\bibinfo {year} {2013})}\BibitemShut {NoStop}%
\bibitem [{ine({\natexlab{b}})}]{ines_gwendal}%
  \BibitemOpen
 \href@noop {} {} \ \bibinfo {note} {R. Bisognin, H. Bartolomei, M. Kumar, I. Safi, J.-M. Berroir, E. Bocquillon, B. Pla\c{c}ais,
  A. Cavanna, U. Gennser, Y. Jin, et G. F\`eve, Nat. Comm. {\bf 10}, 1708
  (2019).}\BibitemShut {Stop}%
\bibitem [{sam()}]{saminad_ines}%
  \BibitemOpen
  \href@noop {} {}\bibinfo {note}
  {\bibinfo{author}{\bibfnamefont{L.}~\bibnamefont{Saminadayar}} {\it et al},
  \bibinfo{journal}{Phys. Rev. Lett.} \textbf{\bibinfo{volume}{79}},
  \bibinfo{pages}{2526} (\bibinfo{year}{1997}).
  \bibinfo{author}{\bibfnamefont{I.}~\bibnamefont{Safi}},
  \bibinfo{journal}{Phys. Rev. B} \textbf{\bibinfo{volume}{55}},
  \bibinfo{pages}{R 12 691} (\bibinfo{year}{1997-II})}\BibitemShut {NoStop}%
\bibitem [{chr()}]{christian_photo_2018}%
  \BibitemOpen
  \href@noop {} {}\bibinfo {note} {M. Kapfer, P. Roulleau, I. Farrer, D. A.
  Ritchie and D. C. Glattli, Science Vol. 363, 846 (2019)}\BibitemShut
  {NoStop}%
\bibitem [{\citenamefont {Safi}\ \emph {et~al.}(2001)\citenamefont {Safi},
  \citenamefont {Devillard},\ and\ \citenamefont {Martin}}]{ines_prl}%
  \BibitemOpen
  \bibfield  {author} {\bibinfo {author} {\bibfnamefont {I.}~\bibnamefont
  {Safi}}, \bibinfo {author} {\bibfnamefont {P.}~\bibnamefont {Devillard}},\
  and\ \bibinfo {author} {\bibfnamefont {T.}~\bibnamefont {Martin}},\
  } {\bibfield  {journal}
  {\bibinfo  {journal} {Phys. Rev. Lett.}\ }\textbf {\bibinfo {volume} {86}},\
  \bibinfo {pages} {4628} (\bibinfo {year} {2001})}\BibitemShut {NoStop}%
\bibitem [{\citenamefont {Kim}\ \emph {et~al.}(2005)\citenamefont {Kim},
  \citenamefont {Lawler}, \citenamefont {Vishveshwara},\ and\ \citenamefont
  {Fradkin}}]{kim_hbt}%
  \BibitemOpen
  \bibfield  {author} {\bibinfo {author} {\bibfnamefont {E.-A.}\ \bibnamefont
  {Kim}}, \bibinfo {author} {\bibfnamefont {M.}~\bibnamefont {Lawler}},
  \bibinfo {author} {\bibfnamefont {S.}~\bibnamefont {Vishveshwara}},\ and\
  \bibinfo {author} {\bibfnamefont {E.}~\bibnamefont {Fradkin}},\ }, {\bibfield  {journal}
  {\bibinfo  {journal} {Physical Review Letters}\ }\textbf {\bibinfo {volume}
  {95}},\ \bibinfo {eid} {176402} (\bibinfo {year} {2005})}\BibitemShut
  {NoStop}%
\bibitem [{ine({\natexlab{c}})}]{ines_schulz_tierce}%
  \BibitemOpen
  \href@noop {} {}\ \bibinfo {note} {I. Safi and H.~J.
  Schulz, in {\em Correlated Fermions and Transport in Mesoscopic Systems}, Eds
  T. Martin, G. Montambaux, and J.~T.~T. Van (Fronti\`eres) (1996); {\it ibid},
  Phys. Rev. B {\bf 59}, 3040 (1999). I. Safi, Eur. Phys. J. B {\bf 12}, 451
  (1999). I.~P. Levkivskyi and E.~V. Sukhorukov, Phys. Rev. B {\bf 78}, 045322
  (2008), \bibinfo{pages}{6854 EP} (\bibinfo{year}{2015})}\BibitemShut
  {NoStop}%
\bibitem [{exp({\natexlab{a}})}]{exp_frac}%
  \BibitemOpen
  \href@noop {} {} \ \bibinfo {note} {H. Kamata {\it et al},
  Nat. Nanotechnol. {\bf 9}, 177 (2014). {Freulon {\it et~al.}},
  \bibinfo{journal}{Nat. Commun.} \textbf{\bibinfo{volume}{6}},
  \bibinfo{pages}{6854 EP} (\bibinfo{year}{2015})}\BibitemShut {NoStop}%
\bibitem [{\citenamefont {Callen}\ and\ \citenamefont
  {Welton}(1951)}]{callen_welton_1951}%
  \BibitemOpen
  \bibfield  {author} {\bibinfo {author} {\bibfnamefont {H.~B.}\ \bibnamefont
  {Callen}}\ and\ \bibinfo {author} {\bibfnamefont {T.~A.}\ \bibnamefont
  {Welton}},\ }
  {\bibfield  {journal} {\bibinfo  {journal} {Phys. Rev.}\ }\textbf {\bibinfo
  {volume} {83}},\ \bibinfo {pages} {34} (\bibinfo {year} {1951})}\BibitemShut
  {NoStop}%
\bibitem [{FDT()}]{FDT_zero}%
  \BibitemOpen
  \href@noop {} {}\bibinfo {note} {{J. Tobiska and Y.~V. Nazarov, Phys. Rev. B
  {\bf 72}, 235328 (2005). K. Saito and Y. Utsumi, Phys. Rev. B {\bf 78},
  115429 (2008). H. F\"orster and M. B\"uttiker, Phys. Rev. Lett. {\bf 101},
  136805 (2008).}}\BibitemShut {Stop}%
  
 \bibitem{hekking_subgap_noise_application_perturbative_approach}
 F. Pistolesi, G. Bignon, and F. W. J. Hekking, Phys. Rev. B {\bf 69}, 214518 (2004).
 
\bibitem [{sup()}]{supplemental}%
  \BibitemOpen
  \href@noop {} {}\bibinfo {note} {See the Supplemental Material.}\BibitemShut
  {Stop}%
\bibitem [{ine({\natexlab{d}})}]{ines_philippe_group}%
  \BibitemOpen
  \href@noop {} {} \ \bibinfo {note} {I. Safi, arxiv-0908.4382
  (2009). I. Safi and P. Joyez, Phys. Rev. B {\bf 84}, 205129
  (2011).}\BibitemShut {Stop}%
\bibitem [{\citenamefont {Bartolomei}\ \emph {et~al.}(2020)\citenamefont
  {Bartolomei}, \citenamefont {Kumar}, \citenamefont {Bisognin}, \citenamefont
  {Marguerite}, \citenamefont {Berroir}, \citenamefont {Bocquillon},
  \citenamefont {Pla{\c c}ais}, \citenamefont {Cavanna}, \citenamefont {Dong},
  \citenamefont {Gennser}, \citenamefont {Jin},\ and\ \citenamefont
  {F{\`e}ve}}]{fractional_statistics_gwendal_science_2020}%
  \BibitemOpen
  \bibfield  {author} {\bibinfo {author} {\bibfnamefont {H.}~\bibnamefont
  {Bartolomei}}, \bibinfo {author} {\bibfnamefont {M.}~\bibnamefont {Kumar}},
  \bibinfo {author} {\bibfnamefont {R.}~\bibnamefont {Bisognin}}, \bibinfo
  {author} {\bibfnamefont {A.}~\bibnamefont {Marguerite}}, \bibinfo {author}
  {\bibfnamefont {J.-M.}\ \bibnamefont {Berroir}}, \bibinfo {author}
  {\bibfnamefont {E.}~\bibnamefont {Bocquillon}}, \bibinfo {author}
  {\bibfnamefont {B.}~\bibnamefont {Pla{\c c}ais}}, \bibinfo {author}
  {\bibfnamefont {A.}~\bibnamefont {Cavanna}}, \bibinfo {author} {\bibfnamefont
  {Q.}~\bibnamefont {Dong}}, \bibinfo {author} {\bibfnamefont {U.}~\bibnamefont
  {Gennser}}, \bibinfo {author} {\bibfnamefont {Y.}~\bibnamefont {Jin}},\ and\
  \bibinfo {author} {\bibfnamefont {G.}~\bibnamefont {F{\`e}ve}},\ } {\bibfield  {journal}
  {\bibinfo  {journal} {Science}\ }\textbf {\bibinfo {volume} {368}},\ \bibinfo
  {pages} {173} (\bibinfo {year} {2020})}\ \BibitemShut
  {NoStop}%
\bibitem [{\citenamefont {Washio}\ \emph {et~al.}(2016)\citenamefont {Washio},
  \citenamefont {Nakazawa}, \citenamefont {Hashisaka}, \citenamefont {Muraki},
  \citenamefont {Tokura},\ and\ \citenamefont
  {Fujisawa}}]{non_thermal_TLL_fujisawa_PRB_2016}%
  \BibitemOpen
  \bibfield  {author} {\bibinfo {author} {\bibfnamefont {K.}~\bibnamefont
  {Washio}}, \bibinfo {author} {\bibfnamefont {R.}~\bibnamefont {Nakazawa}},
  \bibinfo {author} {\bibfnamefont {M.}~\bibnamefont {Hashisaka}}, \bibinfo
  {author} {\bibfnamefont {K.}~\bibnamefont {Muraki}}, \bibinfo {author}
  {\bibfnamefont {Y.}~\bibnamefont {Tokura}},\ and\ \bibinfo {author}
  {\bibfnamefont {T.}~\bibnamefont {Fujisawa}},\ } {\bibfield  {journal} {\bibinfo
  {journal} {Phys. Rev. B}\ }\textbf {\bibinfo {volume} {93}},\ \bibinfo
  {pages} {075304} (\bibinfo {year} {2016})}\BibitemShut {NoStop}%
\bibitem [{the()}]{thermal_group}%
  \BibitemOpen
  \href@noop {} {}\bibinfo {note} {I.~P. Levkivskyi and E. Sukhorukov, Phys.
  Rev. Lett. \textbf{\bibinfo{volume}{109}}, {246806} ({2012}). E. Sivre {\it
  et al}, Nat. Phys. {\bf 14}, 145 (2017). E. S. Tikhonov {\it et al}, arxiv:
  2001.07563}\BibitemShut {NoStop}%
\bibitem [{rev()}]{rev_optics}%
  \BibitemOpen
  \href@noop {} {}\bibinfo {note} {For review, see C. Grenier {\it et al}, Mod.
  Phys. Lett. B {\bf 25}, 1053 (2011), and D. C. Glattli and P. S. Roulleau,
  Phys. Status Sol. {\bf 254}, 1600650 (2017).}\BibitemShut {Stop}%
\bibitem [{\citenamefont {{I.P. Levkivskyi}}\ and\ \citenamefont {{E.V.
  Sukhorukov}}(2009)}]{out_of_equilibrium_bosonisation_eugene_PRL_2009}%
  \BibitemOpen
  \bibfield  {author} {\bibinfo {author} {\bibnamefont {{I.P. Levkivskyi}}}\
  and\ \bibinfo {author} {\bibnamefont {{E.V. Sukhorukov}}},\ }\href@noop {}
  {\bibfield  {journal} {\bibinfo  {journal} {Phys. Rev. Lett.}\ }\textbf
  {\bibinfo {volume} {103}},\ \bibinfo {pages} {036801} (\bibinfo {year}
  {2009})}\BibitemShut {NoStop}%
\bibitem [{\citenamefont {Wen}(1992)}]{wen_review_FQHE_1992}%
  \BibitemOpen
  \bibfield  {author} {\bibinfo {author} {\bibfnamefont {X.-G.}\ \bibnamefont
  {Wen}},\ } {\bibfield  {journal} {\bibinfo
  {journal} {Int. J. Mod. Phys. B}\ }\textbf {\bibinfo {volume} {06}},\
  \bibinfo {pages} {1711} (\bibinfo {year} {1992})}\BibitemShut {NoStop}%
\bibitem [{\citenamefont {Levkivskyi}\ \emph {et~al.}(2009)\citenamefont
  {Levkivskyi}, \citenamefont {Boyarsky}, \citenamefont {Fr\"ohlich},\ and\
  \citenamefont {Sukhorukov}}]{sukho_mac_zhender_PRB_2009}%
  \BibitemOpen
  \bibfield  {author} {\bibinfo {author} {\bibfnamefont {I.~P.}\ \bibnamefont
  {Levkivskyi}}, \bibinfo {author} {\bibfnamefont {A.}~\bibnamefont
  {Boyarsky}}, \bibinfo {author} {\bibfnamefont {J.}~\bibnamefont
  {Fr\"ohlich}},\ and\ \bibinfo {author} {\bibfnamefont {E.~V.}\ \bibnamefont
  {Sukhorukov}},\ } {\bibfield  {journal} {\bibinfo
  {journal} {Phys. Rev. B}\ }\textbf {\bibinfo {volume} {80}},\ \bibinfo
  {pages} {045319} (\bibinfo {year} {2009})}\BibitemShut {NoStop}%
\bibitem [{\citenamefont {Rosenow}\ \emph {et~al.}(2016)\citenamefont
  {Rosenow}, \citenamefont {Levkivskyi},\ and\ \citenamefont
  {Halperin}}]{halperin_HBT_FQHE_2016}%
  \BibitemOpen
  \bibfield  {author} {\bibinfo {author} {\bibfnamefont {B.}~\bibnamefont
  {Rosenow}}, \bibinfo {author} {\bibfnamefont {I.~P.}\ \bibnamefont
  {Levkivskyi}},\ and\ \bibinfo {author} {\bibfnamefont {B.~I.}\ \bibnamefont
  {Halperin}},\ } {\bibfield  {journal}
  {\bibinfo  {journal} {Phys. Rev. Lett.}\ }\textbf {\bibinfo {volume} {116}},\
  \bibinfo {pages} {156802} (\bibinfo {year} {2016})}\BibitemShut {NoStop}%
\bibitem [{\citenamefont {Safi}\ and\ \citenamefont
  {Sukhorukov}(2010)}]{ines_eugene}%
  \BibitemOpen
  \bibfield  {author} {\bibinfo {author} {\bibfnamefont {I.}~\bibnamefont
  {Safi}}\ and\ \bibinfo {author} {\bibfnamefont {E.~V.}\ \bibnamefont
  {Sukhorukov}},\ } {\bibfield  {journal}
  {\bibinfo  {journal} {Eur. Phys. Lett.}\ }\textbf {\bibinfo {volume} {91}},\
  \bibinfo {pages} {67008} (\bibinfo {year} {2010})}\BibitemShut {NoStop}%
\bibitem [{exp({\natexlab{b}})}]{expt_symm}%
  \BibitemOpen
  \href@noop {} {} \ \bibinfo {note} {E. Zakka-Bajjani {\it
  et~al.}, Phys. Rev. Lett. {\bf 99}, 236803 (2007); J. Gabelli and B. Reulet,
  Phys. Rev. Lett. {\bf 100}, 026601 (2008)}\BibitemShut {NoStop}%
\bibitem [{exc()}]{excess}%
  \BibitemOpen
  \href@noop {} {}\bibinfo {note} {To deduce the excess noise,
  $S(\omega_J;\omega)-S_{eq}(\omega)$, one cannot infer the equilibrium noise
  $S_{eq}(\omega)$ from $S(\omega_J=0;\omega)$, as all additional dc drives
  must vanish too. Notice that Eq. \eqref{noise_DC_initial_sym_0} is different from the Rogovin-Scalapino's FDR written in terms of the equilibrium noise with translated frequency arguments}\BibitemShut {Stop}%
\bibitem [{adm()}]{admittance_definition}%
  \BibitemOpen
  \href@noop {} {}\bibinfo {note} {${Y}(\omega_J,\omega)$ is the response of
  average current to a small ac modulation $\delta V(t)= v_{ac}e^{i \omega t}$
  superimposed on $V_{dc}.$ In particular,
  ${Y}(\omega_J,\omega=0)=G_{dc}(\omega_J)$}\BibitemShut {NoStop}%
\bibitem [{\citenamefont {Altimiras}\ \emph {et~al.}(2010)\citenamefont
  {Altimiras}, \citenamefont {le~Sueur}, \citenamefont {Gennser}, \citenamefont
  {Cavanna}, \citenamefont {Mailly},\ and\ \citenamefont
  {Pierre}}]{pierre_equilibration_IQHE_Nature_2010}%
  \BibitemOpen
  \bibfield  {author} {\bibinfo {author} {\bibfnamefont {C.}~\bibnamefont
  {Altimiras}}, \bibinfo {author} {\bibfnamefont {H.}~\bibnamefont {le~Sueur}},
  \bibinfo {author} {\bibfnamefont {U.}~\bibnamefont {Gennser}}, \bibinfo
  {author} {\bibfnamefont {A.}~\bibnamefont {Cavanna}}, \bibinfo {author}
  {\bibfnamefont {D.}~\bibnamefont {Mailly}},\ and\ \bibinfo {author}
  {\bibfnamefont {F.}~\bibnamefont {Pierre}},\ } {\bibfield
  {journal} {\bibinfo  {journal} {Nat. Phys.}\ }\textbf {\bibinfo {volume}
  {6}},\ \bibinfo {pages} {34} (\bibinfo {year} {2010})}\BibitemShut {NoStop}%
\bibitem [{ine({\natexlab{e}})}]{ines_average}%
  \BibitemOpen
  \href@noop \ \bibinfo {note} {I. Safi, unpublished
  (2020).}\BibitemShut {Stop}%
\bibitem [{\citenamefont {Di~Marco}\ \emph {et~al.}(2013)\citenamefont
  {Di~Marco}, \citenamefont {Maisi}, \citenamefont {Pekola},\ and\
  \citenamefont {Hekking}}]{hekking_various_T_PRB_2013}%
  \BibitemOpen
  \bibfield  {author} {\bibinfo {author} {\bibfnamefont {A.}~\bibnamefont
  {Di~Marco}}, \bibinfo {author} {\bibfnamefont {V.~F.}\ \bibnamefont {Maisi}},
  \bibinfo {author} {\bibfnamefont {J.~P.}\ \bibnamefont {Pekola}},\ and\
  \bibinfo {author} {\bibfnamefont {F.~W.~J.}\ \bibnamefont {Hekking}},\
  },  {\bibfield  {journal} {\bibinfo
  {journal} {Phys. Rev. B}\ }\textbf {\bibinfo {volume} {88}},\ \bibinfo
  {pages} {174507} (\bibinfo {year} {2013})}\BibitemShut {NoStop}%
  \bibitem [{\citenamefont {Dolcetto}\ \emph {et~al.}(2014)\citenamefont
  {Dolcetto}, \citenamefont {Cavaliere},\ and\ \citenamefont
  {Sassetti}}]{sassetti_spin_hall_LL_PRB_2014}%
  \BibitemOpen
  \bibfield  {author} {\bibinfo {author} {\bibfnamefont {G.}~\bibnamefont
  {Dolcetto}}, \bibinfo {author} {\bibfnamefont {F.}~\bibnamefont
  {Cavaliere}},\ and\ \bibinfo {author} {\bibfnamefont {M.}~\bibnamefont
  {Sassetti}},\ } {\bibfield  {journal} {\bibinfo
  {journal} {Phys. Rev. B}\ }\textbf {\bibinfo {volume} {89}},\ \bibinfo
  {pages} {125419} (\bibinfo {year} {2014})}\BibitemShut {NoStop}%
  \bibitem{spin_hall_glazman}Jukka I. Vayrynen and Leonid I. Glazman,
Phys. Rev. Lett. {\bf{118}}, 106802 (2017).

  
\end{thebibliography}

\begin{thebibliography}{18}%
\makeatletter
\providecommand \@ifxundefined [1]{%
 \@ifx{#1\undefined}
}%
\providecommand \@ifnum [1]{%
 \ifnum #1\expandafter \@firstoftwo
 \else \expandafter \@secondoftwo
 \fi
}%
\providecommand \@ifx [1]{%
 \ifx #1\expandafter \@firstoftwo
 \else \expandafter \@secondoftwo
 \fi
}%
\providecommand \natexlab [1]{#1}%
\providecommand \enquote  [1]{``#1''}%
\providecommand \bibnamefont  [1]{#1}%
\providecommand \bibfnamefont [1]{#1}%
\providecommand \citenamefont [1]{#1}%
\providecommand \href@noop [0]{\@secondoftwo}%
\providecommand \href [0]{\begingroup \@sanitize@url \@href}%
\providecommand \@href[1]{\@@startlink{#1}\@@href}%
\providecommand \@@href[1]{\endgroup#1\@@endlink}%
\providecommand \@sanitize@url [0]{\catcode `\\12\catcode `\$12\catcode
  `\&12\catcode `\#12\catcode `\^12\catcode `\_12\catcode `\%12\relax}%
\providecommand \@@startlink[1]{}%
\providecommand \@@endlink[0]{}%
\providecommand \url  [0]{\begingroup\@sanitize@url \@url }%
\providecommand \@url [1]{\endgroup\@href {#1}{\urlprefix }}%
\providecommand \urlprefix  [0]{URL }%
\providecommand \Eprint [0]{\href }%
\providecommand \doibase [0]{https://doi.org/}%
\providecommand \selectlanguage [0]{\@gobble}%
\providecommand \bibinfo  [0]{\@secondoftwo}%
\providecommand \bibfield  [0]{\@secondoftwo}%
\providecommand \translation [1]{[#1]}%
\providecommand \BibitemOpen [0]{}%
\providecommand \bibitemStop [0]{}%
\providecommand \bibitemNoStop [0]{.\EOS\space}%
\providecommand \EOS [0]{\spacefactor3000\relax}%
\providecommand \BibitemShut  [1]{\csname bibitem#1\endcsname}%
\let\auto@bib@innerbib\@empty
\bibitem [{\citenamefont {Davids}\ and\ \citenamefont
  {Shank}(2018)}]{tunneling_non_perturbative_PRB_2018}%
  \BibitemOpen
  \bibfield  {author} {\bibinfo {author} {\bibfnamefont {P.~S.}\ \bibnamefont
  {Davids}}\ and\ \bibinfo {author} {\bibfnamefont {J.}~\bibnamefont {Shank}},\
  }\bibfield  {title} {\bibinfo {title} {Density matrix approach to
  photon-assisted tunneling in the transfer hamiltonian formalism},\ }\href
  {https://doi.org/10.1103/PhysRevB.97.075411} {\bibfield  {journal} {\bibinfo
  {journal} {Phys. Rev. B}\ }\textbf {\bibinfo {volume} {97}},\ \bibinfo
  {pages} {075411} (\bibinfo {year} {2018})}\BibitemShut {NoStop}%
\bibitem [{ine({\natexlab{a}})}]{ines_cond_mat_prb}%
  \BibitemOpen
  \href@noop {} {} \ \bibinfo {note} {I. Safi, arxiv:1401.5950
  (2014). Phys. Rev. B {\bf{99}}, 045101 (2019).}\BibitemShut {Stop}%
\bibitem [{\citenamefont {Safi}\ and\ \citenamefont
  {Schulz}(1999)}]{ines_prb_long}%
  \BibitemOpen
  \bibfield  {author} {\bibinfo {author} {\bibfnamefont {I.}~\bibnamefont
  {Safi}}\ and\ \bibinfo {author} {\bibfnamefont {H.~J.}\ \bibnamefont
  {Schulz}},\ }\bibfield  {title} {\bibinfo {title} {Interacting electrons with
  spin in a one-dimensional dirty wire connected to leads},\ }\href@noop {}
  {\bibfield  {journal} {\bibinfo  {journal} {Phys. Rev. B}\ }\textbf {\bibinfo
  {volume} {59}},\ \bibinfo {pages} {3040} (\bibinfo {year}
  {1999})}\BibitemShut {NoStop}%
\bibitem [{\citenamefont {Roussel}\ \emph {et~al.}(2016)\citenamefont
  {Roussel}, \citenamefont {Degiovanni},\ and\ \citenamefont
  {Safi}}]{ines_degiovanni_2016}%
  \BibitemOpen
  \bibfield  {author} {\bibinfo {author} {\bibfnamefont {B.}~\bibnamefont
  {Roussel}}, \bibinfo {author} {\bibfnamefont {P.}~\bibnamefont
  {Degiovanni}},\ and\ \bibinfo {author} {\bibfnamefont {I.}~\bibnamefont
  {Safi}},\ }\bibfield  {title} {\bibinfo {title} {Perturbative fluctuation
  dissipation relation for nonequilibrium finite-frequency noise in quantum
  circuits},\ } {\bibfield
  {journal} {\bibinfo  {journal} {Phys. Rev. B}\ }\textbf {\bibinfo {volume}
  {93}},\ \bibinfo {pages} {045102} (\bibinfo {year} {2016})}\BibitemShut
  {NoStop}%
\bibitem [{ben()}]{bena}%
  \BibitemOpen
  \href@noop {} {}\bibinfo {note} {C. Bena and I. Safi, Phys. Rev. B {\bf 76},
  125317 (2007). I. Safi, C. Bena, and A. Cr\'epieux, Phys. Rev. B {\bf 78},
  205422 (2008).}\BibitemShut {Stop}%
\bibitem [{zam()}]{zamoum_souquet}%
  \BibitemOpen
  \href@noop {} {}\bibinfo {note} {R. Zamoum, A. Cr\'epieux, and I. Safi, Phys.
  Rev. B {\bf 85}, 125421 (2012). J.-R. Souquet, I. Safi, and P. Simon, Phys.
  Rev. B {\bf 88}, 205419 (2013).}\BibitemShut {Stop}%
\bibitem [{\citenamefont {Rogovin}\ and\ \citenamefont
  {Scalapino}(1974)}]{FDT_rogovin_scalapino_74}%
  \BibitemOpen
  \bibfield  {author} {\bibinfo {author} {\bibfnamefont {D.}~\bibnamefont
  {Rogovin}}\ and\ \bibinfo {author} {\bibfnamefont {D.}~\bibnamefont
  {Scalapino}},\ }\bibfield  {title} {\bibinfo {title} {Fluctuation phenomena
  in tunnel junctions},\ } {\bibfield
  {journal} {\bibinfo  {journal} {Annals of Physics}\ }\textbf {\bibinfo
  {volume} {86}},\ \bibinfo {pages} {1 } (\bibinfo {year} {1974})}\BibitemShut
  {NoStop}%
\bibitem [{any()}]{anyon_collider}%
  \BibitemOpen
  \href@noop {} {}\bibinfo {note}
  {\bibinfo{author}{\bibfnamefont{B.}~\bibnamefont{Rosenow}},
  \bibinfo{author}{\bibfnamefont{I.~P.} \bibnamefont{Levkivskyi}},
  \bibnamefont{and} \bibinfo{author}{\bibfnamefont{B.~I.}
  \bibnamefont{Halperin}}, \bibinfo{journal}{Phys. Rev. Lett.}
  \textbf{\bibinfo{volume}{116}}, \bibinfo{pages}{156802}
  (\bibinfo{year}{2016}).}\BibitemShut {Stop}%
\bibitem [{\citenamefont {Bartolomei}\ \emph {et~al.}(2020)\citenamefont
  {Bartolomei}, \citenamefont {Kumar}, \citenamefont {Bisognin}, \citenamefont
  {Marguerite}, \citenamefont {Berroir}, \citenamefont {Bocquillon},
  \citenamefont {Pla{\c c}ais}, \citenamefont {Cavanna}, \citenamefont {Dong},
  \citenamefont {Gennser}, \citenamefont {Jin},\ and\ \citenamefont
  {F{\`e}ve}}]{fractional_statistics_gwendal_science_2020}%
  \BibitemOpen
  \bibfield  {author} {\bibinfo {author} {\bibfnamefont {H.}~\bibnamefont
  {Bartolomei}}, \bibinfo {author} {\bibfnamefont {M.}~\bibnamefont {Kumar}},
  \bibinfo {author} {\bibfnamefont {R.}~\bibnamefont {Bisognin}}, \bibinfo
  {author} {\bibfnamefont {A.}~\bibnamefont {Marguerite}}, \bibinfo {author}
  {\bibfnamefont {J.-M.}\ \bibnamefont {Berroir}}, \bibinfo {author}
  {\bibfnamefont {E.}~\bibnamefont {Bocquillon}}, \bibinfo {author}
  {\bibfnamefont {B.}~\bibnamefont {Pla{\c c}ais}}, \bibinfo {author}
  {\bibfnamefont {A.}~\bibnamefont {Cavanna}}, \bibinfo {author} {\bibfnamefont
  {Q.}~\bibnamefont {Dong}}, \bibinfo {author} {\bibfnamefont {U.}~\bibnamefont
  {Gennser}}, \bibinfo {author} {\bibfnamefont {Y.}~\bibnamefont {Jin}},\ and\
  \bibinfo {author} {\bibfnamefont {G.}~\bibnamefont {F{\`e}ve}},\ }\bibfield
  {title} {\bibinfo {title} {Fractional statistics in anyon collisions},\
  }\href {https://doi.org/10.1126/science.aaz5601} {\bibfield  {journal}
  {\bibinfo  {journal} {Science}\ }\textbf {\bibinfo {volume} {368}},\ \bibinfo
  {pages} {173} (\bibinfo {year} {2020})},\  \BibitemShut
  {NoStop}%
    \bibitem [{\citenamefont {Kane}\ and\ \citenamefont
  {Fisher}(2003)}]{kane_fisher_dilute}%
  \BibitemOpen
  \bibfield  {author} {\bibinfo {author} {\bibfnamefont {C.~L.}\ \bibnamefont
  {Kane}}\ and\ \bibinfo {author} {\bibfnamefont {M.~P.~A.}\ \bibnamefont
  {Fisher}},\ }\bibfield  {title} {\bibinfo {title} {Shot noise and the
  transmission of dilute laughlin quasiparticles},\ }\href
  {https://doi.org/10.1103/PhysRevB.67.045307} {\bibfield  {journal} {\bibinfo
  {journal} {Phys. Rev. B}\ }\textbf {\bibinfo {volume} {67}},\ \bibinfo
  {pages} {045307} (\bibinfo {year} {2003})}\BibitemShut {NoStop}%
\bibitem [{ine({\natexlab{c}})}]{ines_cond_09}%
  \BibitemOpen
  \href@noop {} {} \ \bibinfo {note} {I. Safi,
  arxiv:0908.4382009 (2009).}\BibitemShut {Stop}%
\bibitem [{\citenamefont {{I.P. Levkivskyi}}\ and\ \citenamefont {{E.V.
  Sukhorukov}}(2009)}]{out_of_equilibrium_bosonisation_eugene_PRL_2009}%
  \BibitemOpen
  \bibfield  {author} {\bibinfo {author} {\bibnamefont {{I.P. Levkivskyi}}}\
  and\ \bibinfo {author} {\bibnamefont {{E.V. Sukhorukov}}},\ }\href@noop {}
  {\bibfield  {journal} {\bibinfo  {journal} {Phys. Rev. Lett.}\ }\textbf
  {\bibinfo {volume} {103}},\ \bibinfo {pages} {036801} (\bibinfo {year}
  {2009})}\BibitemShut {NoStop}%
\bibitem [{\citenamefont {Safi}(1999)}]{ines_epj}%
  \BibitemOpen
  \bibfield  {author} {\bibinfo {author} {\bibfnamefont {I.}~\bibnamefont
  {Safi}},\ }\bibfield  {title} {\bibinfo {title} {A dynamic scattering
  approach for a gated interacting wire},\ }\href@noop {} {\bibfield  {journal}
  {\bibinfo  {journal} {Eur. Phys. J. B}\ }\textbf {\bibinfo {volume} {12}},\
  \bibinfo {pages} {451} (\bibinfo {year} {1999})}\BibitemShut {NoStop}%
\bibitem [{not()}]{note_OE}%
  \BibitemOpen
  \href@noop {} {}\bibinfo {note} {Notice that the OE formalism we have
  developed in Ref.\cite{bena,trauzettel_group_tierce} is a form of OE
  bosonisation; it takes into account the effect of the second cumulant at a
  QPC, thus its high frequency noise, on the bosonic Green's functions
  evaluated at an arbitrary distance from that QPC. This could be sufficient
  only if higher cumulants can be ignored, as in the IQHE with a plasmonic
  dispersion \cite{thermal_eugene_Hall_PRL_2012}.}\BibitemShut {Stop}%
\bibitem [{tra()}]{trauzettel_group_tierce}%
  \BibitemOpen
  \href@noop {} {}\bibinfo {note} {{B. Trauzettel}, {I. Safi}, {F. Dolcini},
  and {H. Grabert}, Phys. Rev. Lett. {\bf 92}, 226405 (2004). F. Dolcini, B.
  Trauzettel, I. Safi, and H. Grabert, Phys. Rev. B {\bf 71}, 165309
  (2005).}\BibitemShut {Stop}%
\bibitem [{\citenamefont {Levkivskyi}\ and\ \citenamefont
  {Sukhorukov}(2012)}]{thermal_eugene_Hall_PRL_2012}%
  \BibitemOpen
  \bibfield  {author} {\bibinfo {author} {\bibfnamefont {I.~P.}\ \bibnamefont
  {Levkivskyi}}\ and\ \bibinfo {author} {\bibfnamefont {E.~V.}\ \bibnamefont
  {Sukhorukov}},\ } {\bibfield  {journal}
  {\bibinfo  {journal} {Phys. Rev. Lett.}\ }\textbf {\bibinfo {volume} {109}},\
  \bibinfo {pages} {246806} (\bibinfo {year} {2012})}\BibitemShut {NoStop}%
\end{thebibliography}
\end{document}